\documentclass[11pt,letterpaper]{article}

\usepackage[letterpaper,margin=1.0in]{geometry}
\usepackage{url}
\usepackage{amssymb}
\usepackage{graphicx}
\usepackage{amsmath}
\usepackage{amsfonts}
\usepackage{amssymb}
\usepackage{amsthm}
\usepackage{multirow}
\usepackage{algorithm}
\usepackage{algorithmic}

\usepackage[pdftex,
            colorlinks,
            plainpages=false,
            hyperindex,
            bookmarksopen,
            linkcolor=black,
            citecolor=black,
            urlcolor=blue]{hyperref}
\usepackage[labelfont=bf]{subcaption}
\usepackage{multicol}
\usepackage{multirow}
\usepackage{xcolor}
\usepackage{float}

\usepackage{authblk}

\usepackage[normalem]{ulem}

\newcommand{\KET}[1]{\left| #1\right\rangle }
\newcommand{\MULTIREF}[3]{#1~\ref{#2} to \ref{#3}}
\newcommand{\REF}[2]{#1~\ref{#2}}

\title{Deterministic Algorithms for Compiling Quantum Circuits with Recurrent Patterns}

\author[1,2]{Davide Ferrari}
\author[3]{Ivano Tavernelli}
\author[1,2]{Michele Amoretti\footnote{michele.amoretti@unipr.it}}
\affil[1]{Department of Engineering and Architecture - University of Parma, Italy}
\affil[2]{Quantum Information Science @ University of Parma, Italy}
\affil[3]{IBM Research - Zurich, R\"uschlikon, Switzerland}

\date{}

\begin{document}

\maketitle

\begin{abstract}
Current quantum processors are noisy, have limited coherence and imperfect gate implementations. On such hardware, only algorithms that are shorter than the overall coherence time can be implemented and executed successfully. A good quantum compiler must translate an input program into the most efficient equivalent of itself, getting the most out of the available hardware. In this work, we present novel deterministic algorithms for compiling recurrent quantum circuit patterns in polynomial time. In particular, such patterns appear in quantum circuits that are used to compute the ground state properties of molecular systems using the variational quantum eigensolver (VQE) method together with the RyRz heuristic wavefunction Ans\"atz. 
We show that our pattern-oriented compiling algorithms, combined with an efficient swapping strategy, produces -- in general -- output programs that are comparable to those obtained with state-of-art compilers, in terms of CNOT count and CNOT depth. In particular, our solution produces unmatched results on RyRz circuits.

\textbf{keywords} - \textit{Quantum compilation, Recurrent patterns, RyRz circuits}
\end{abstract}

\section{Introduction}
\label{sec:introduction}

The idea of quantum computing arose in 1982 during a speech by Richard Feynman about the difficulty of simulating quantum mechanical systems with classical computers~\cite{Feynman1982}. Feynman suggested the simulation of these systems using quantum computers, i.e., controlled quantum mechanical systems able to mimic them. Since then, quantum computing and quantum information theory continued to advance, proving that universal quantum computers could become, for some applications, more powerful than Turing machines~\cite{Shor1994}. Quantum computing will potentially have a deep impact on a variety of fields, from quantum simulation in physics and chemistry~\cite{Chiesa2018}, to machine learning~\cite{Lloyd2014,Biamonte2017,Havlicek2019,Zoufal2019,Cong2019}, artificial intelligence~\cite{Tacchino2018} and cryptography~\cite{Ekert1991,Portmann2014,Fitzsimons2017}.

Current quantum computers are noisy, characterized by a reduced number of qubits (5-50) with non-uniform quality and highly constrained connectivity. Such devices may be able to perform tasks which surpass the capabilities of today's most powerful classical digital computers, but noise in quantum gates limits the size of quantum circuits that can be executed reliably.

\subsection{Quantum Compilation Problem}
\label{sec:problem}

The problem of \textit{quantum compilation}, i.e., device-aware implementation of quantum algorithms, is a challenging one. A good quantum compiler must translate an input program into the most efficient equivalent of itself~\cite{Corcoles2020}, getting the most out of the available hardware. In general, the quantum compilation problem is NP-Hard~\cite{Botea2018,Soeken2019}. On noisy devices, quantum compilation is declined in the following tasks: gate synthesis~\cite{Kliuchnikov2016}, which is the decomposition of an arbitrary unitary operation into a quantum circuit made of single-qubit and two-qubit gates from a universal gate set; compliance with the hardware architecture, starting from an initial mapping of the virtual qubits to the physical ones, and moving through subsequent mappings by means of a clever swapping strategy; and noise awareness.
Quality indicators of the compiled quantum algorithm are, for example, circuit depth, number of gates and fidelity of quantum states~\cite{MunozCoreas2019}.

As an example of compliance with the hardware architecture, consider the problem of compiling the circuit in \REF{Fig.}{fig:map_ex_circuit1} onto a device with a coupling map such as the one shown in \REF{Fig.}{fig:map_ex_coupling}. One could choose a trivial initial mapping of virtual qubits to the physical ones, such as the one depicted in red to the left of the circuit in \REF{Fig.}{fig:map_ex_circuit1}. However, with such a mapping, the CNOT between qubits 0 and 3 could not be directly executed on the device. A more suitable mapping is instead the one shown in blue to the left of the circuit in \REF{Fig.}{fig:map_ex_circuit2}, as it enables the execution of all CNOTs onto the device at the cost of inserting only one SWAP gate.

\begin{figure}
\hspace*{\fill}%
\subcaptionbox{\label{fig:map_ex_coupling}}{\includegraphics[width=.2\linewidth]{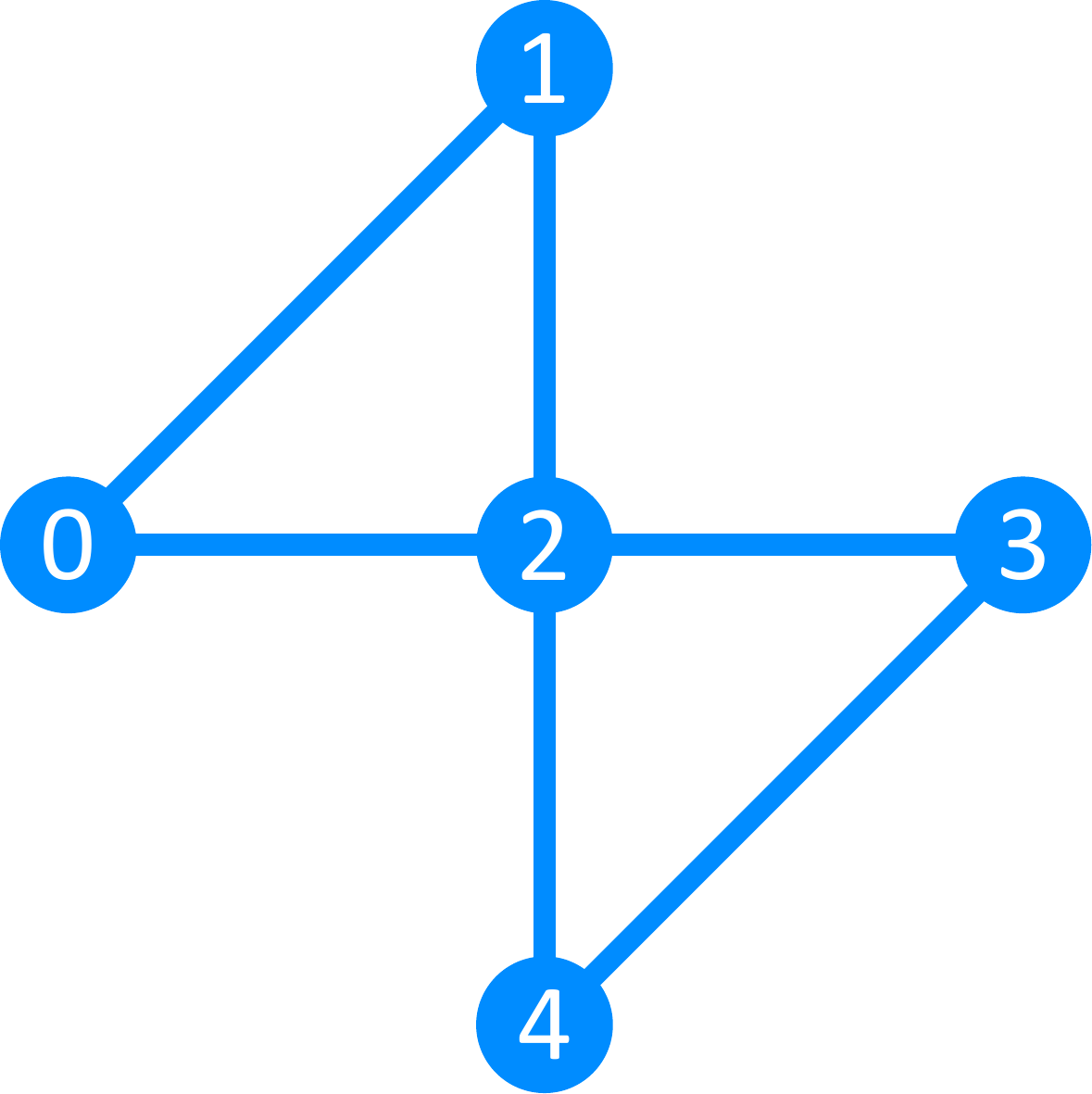}}\hfill%
  \subcaptionbox{\label{fig:map_ex_circuit1}}{\includegraphics[width=.4\linewidth]{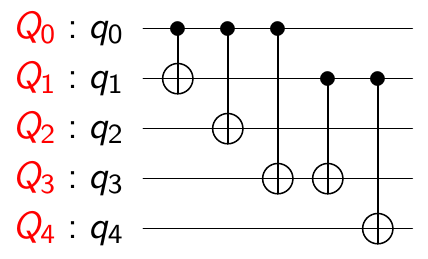}}\hfill%
  \subcaptionbox{\label{fig:map_ex_circuit2}}{\includegraphics[width=.4\linewidth]{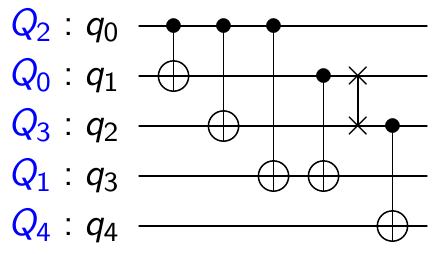}}\hspace*{\fill}%
    \caption{Compiling the 5 qubit circuit shown in \textbf{(b)} onto a 5 qubit device, namely \textit{ibmq\_yorktown}, whose coupling map is depicted in \textbf{(a)}. The compiled circuit is shown in \textbf{(c)}.}
    \label{fig:mapping_example}
\end{figure}

\subsection{Our Contributions}
\label{sec:contributions}
In this paper, we present novel deterministic algorithms for compiling recurrent quantum circuit patterns in polynomial time. In particular, such patterns appear in quantum circuits that are used to compute the ground state properties of molecular systems using the variational quantum eigensolver (VQE) method together with the RyRz heuristic wavefunction Ans\"atz~\cite{Kandala2017}. We implemented our algorithms in a Python software denoted as PADQC (PAttern-oriented Deterministic Quantum Compiler) that we integrated with Qiskit's SABRE swapping strategy~\cite{Li2019} and compilation routine~\cite{QiskitSDK_short}, from now on denoted as Qiskit(SABRE).

We benchmarked PADQC+Qiskit(SABRE) using different quantum circuits, assuming IBM Quantum hardware. We show that our integrated solution produces -- in general -- output programs that are comparable to those obtained with state-of-art compilers, such as t$|$ket$\rangle$~\cite{Sivarajah2020}, in terms of CNOT count and CNOT depth. In particular, our solution produces unmatched results on RyRz circuits.

The paper is organized as follows. In Section \ref{sec:related}, we discuss the state of the art in quantum compiling. In Section \ref{sec:algorithms}, we present our algorithms. In Section \ref{sec:results}, we illustrate the experimental evaluation of PADQC+Qiskit(SABRE). Finally, in Section \ref{sec:conclusions}, we conclude the paper with an outline of future work.

\section{Related work}
\label{sec:related}

Recently, some noteworthy quantum compiling techniques have been proposed. Here we survey those that have been implemented into actual compilers, and benchmarked.

The approach proposed by Zulehner \textit{et al.}~\cite{Zulehner2019} is to partition the circuit into layers, each layer including gates that can be executed in parallel. For each layer, a compliant CNOT mapping must be found, starting from an initial mapping obtained from the previous layer. Denoting the number of physical qubits as $m$ and the number of logical qubits as $n$, in the worst case there are $m!(m-n)!$ possible mappings. Such a huge search space cannot be explored exhaustively. The $A^*$ search algorithm is adopted, to find the less expensive swap sequence. Moreover, a lookahead strategy is adopted to minimize additional operations to switch between subsequent mappings. The proposed solution is efficient in terms of running time and output depth, but may not be scalable because of the exponential space complexity of the $A^*$ search algorithm~\cite{RussellNorvig2020}.

SABRE by Li \textit{et al.}~\cite{Li2019} is a SWAP-based bidirectional heuristic search method. It requires a preprocessing phase consisting of the following steps. First of all, the distance matrix over the coupling map is computed. Then, the directed acyclic graph that represents the two-qubit gate dependencies of the circuit is generated. A data structure denoted as $F$ (front layer) is initialized as the set of two-qubit gates without unexecuted predecessors. The preprocessing phase ends up with the generation of a random initial mapping. Then, the compiling phase consists in iterating the following steps over $F$, until $F$ is empty. First, all executable gates are removed from $F$ and their successors are added to $F$. Second, for those gates in $F$ that cannot be executed, the best SWAP sequence is selected using an heuristic cost function based on distance matrix. Experiment results show that SABRE can generate hardware-compliant circuits with less or comparable overhead, with respect to the approach proposed by Zulehner \textit{et al.}~\cite{Zulehner2019}. 

In Qiskit (version 0.20)~\cite{QiskitSDK_short}, the compiling process is implemented by a customizable Pass Manager that schedules a number of different passes: layout selection, unrolling (i.e., gate synthesis), swap, gate optimization, and more. Four swap strategies are currently available: Basic, Stochastic, Lookahead and SABRE. The Stochastic strategy uses a randomized algorithm to map the input circuit to the selected coupling map. This means that a single run does not guarantee to produce the best result.

Currently, the most advanced quantum compiler is t$|$ket$\rangle$~\cite{Sivarajah2020}, which is written in C++. The compiling process proceeds in two phases: an architecture-independent optimisation phase, which aims to reduce the size and complexity of the circuit; and an architecture-dependent phase, which prepares the circuit for execution on the target machine. The architecture-independent optimisation phase consists of peephole optimizations (targeting small circuit patterns) and macroscopic optimizations (aiming to identify high-level macroscopic structures in the circuit).
The end product of this process is a circuit that can be scheduled for execution by the runtime environment, or simply saved for later.

In Section \ref{sec:results}, we compare our PADQC+Qiskit(SABRE) integration with pure Qiskit(SABRE) and t$|$ket$\rangle$.

\section{Algorithms}
\label{sec:algorithms}

Most compiling approaches aim at finding a general purpose compiler, able to cope with any circuit without the possibility to make assumptions on its structure or characteristics. This kind of solution, although effective in many cases, may not be as much efficient when facing circuits characterized by well-defined peculiar sequences, i.e., \textit{patterns}, of two-qubit operators. This is particularly true if those patterns repeat themselves many times in a circuit and are not compliant with the quantum device connectivity.

This is the case of RyRz circuits used to compute the ground state properties of molecular systems with the variational quantum eigensolver (VQE) method.
These circuits were introduced for the first time in~\cite{Kandala2017} as a heuristic hardware-efficient wavefunction Ans\"atz for the calculation of the electronic structure properties of small molecular systems such as hydrogen H$_{2}$, lithium hydride, LiH, and berillium hydride, BH$_2$, on a quantum computer. 
Contrary to other quantum circuits inspired by classical wavefunction expansion techniques (e.g., the coupled cluster expansion~\cite{Peruzzo2014,Barkoutsos2018}), in this case the nature of the circuit is solely motivated by the requirement of producing an entangled wavefunction for the many-electron systems that optimally fits the connectivity of the hardware at disposal. In most cases, the RyRz circuit offers a well balanced compromise between these two requirements.
These circuits, when implemented with full entanglement (\REF{Fig.}{fig:ryrz_circuits}), are characterized by repeated sequences of a pattern that we denote as \textit{inverted CNOT cascade}.

\begin{figure}
\centering
        \begin{minipage}{10cm}
        \centering
        \includegraphics[width=10cm]{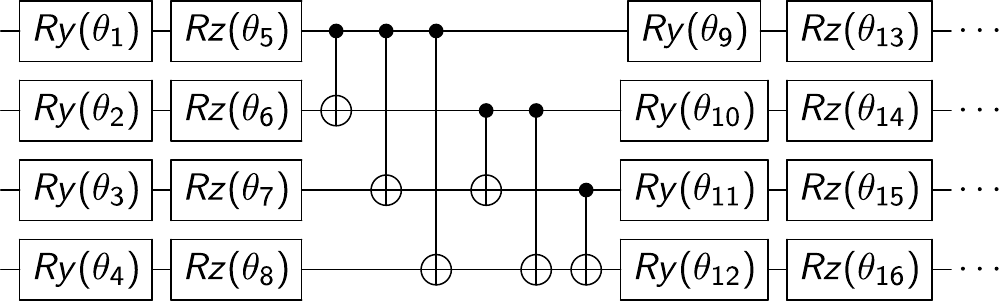}
        \subcaption{}
        \end{minipage}
\caption{RyRz circuit example.}
\label{fig:ryrz_circuits}
\end{figure}

In Sections~\ref{sec:inverted_cascade} and \ref{sec:nn_cnot}, we illustrate these patterns and how their features can be exploited to lay out efficient compiling algorithms. From now on, it will be assumed that the connectivity of the quantum device on which the circuit has to be compiled is similar to those featured by IBM Quantum devices~\cite{new_backends} (\REF{Fig.}{fig:coupling_maps}). \REF{Table}{tab:algos} presents an overview of the designed algorithms and their time complexity, which is computed by taking into account the worst case scenario and the running time of the subroutines.

The compilation process starts with \REF{Algorithm}{alg:patterns}, which searches for patterns of interests and transforms them so that they are more easily mappable to the coupling map. Then, \REF{Algorithm}{alg:gate_cancellation} optimizes the circuit, removing double CNOTs and double \textit{H} gates that may results from the previous transformations. Finally, \REF{Algorithm}{alg:chain} finds a suitable initial mapping for the circuit.

\begin{table}
    \centering
    \resizebox{15cm}{!}{
    {\begin{tabular}{l|c|c}
        \textbf{Algorithm} & \textbf{Subroutines} & \textbf{Time Complexity} \\\hline
        \multirow{2}{*}{\textbf{1} \textsc{patterns}} & \textsc{CheckCascade} & \multirow{2}{*}{$O(g)$}\\
        &\textsc{CheckInverseCascade}\\\hline
        \textbf{2} \textsc{CheckCascade} &  & $O(m)$\\\hline
        \textbf{3} \textsc{CnotCancellation} &  & $O(lm^2)$\\\hline
        \textbf{4} \textsc{GateCancellation} &  & $O(lm^2)$\\\hline
        \multirow{2}{*}{\textbf{5} \textsc{Chain}} & \textsc{CheckForIsolated} & \multirow{2}{*}{$O(n)$}\\
        &\textsc{ExpandChain}\\\hline
        \textbf{6} \textsc{CheckForIsolated} &  & $O(1)$\\\hline
        \textbf{7} \textsc{ExpandChain} &  & $O(1)$\\
    \end{tabular}}}
    \caption{Overview of proposed algorithms and their time complexity. Notation: $n$ is the number of qubits of the device, $m$ is the number of qubits used by the compiled circuit, $g$ is the number of gates in the circuit and $l$ the number of layers.}
    \label{tab:algos}
\end{table}

\begin{figure}
\centering
	\begin{tabular}{ cc }
        \begin{minipage}{5cm}
        \centering
		\includegraphics[width=5cm]{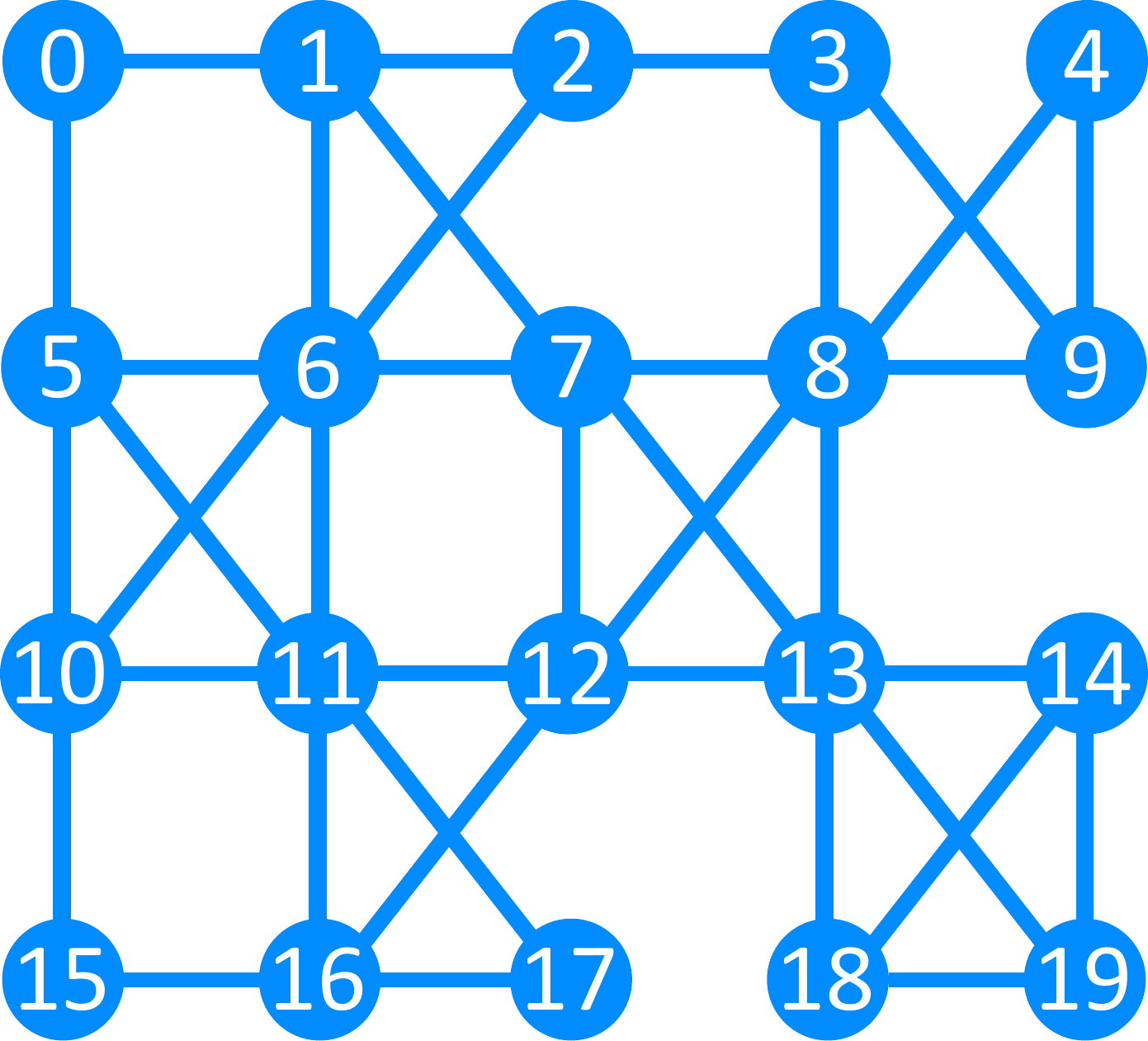}
		\subcaption{}
		\label{fig:coupling_maps_a}
		\end{minipage} &
		\begin{minipage}{5cm}
		\centering
		\includegraphics[width=5cm]{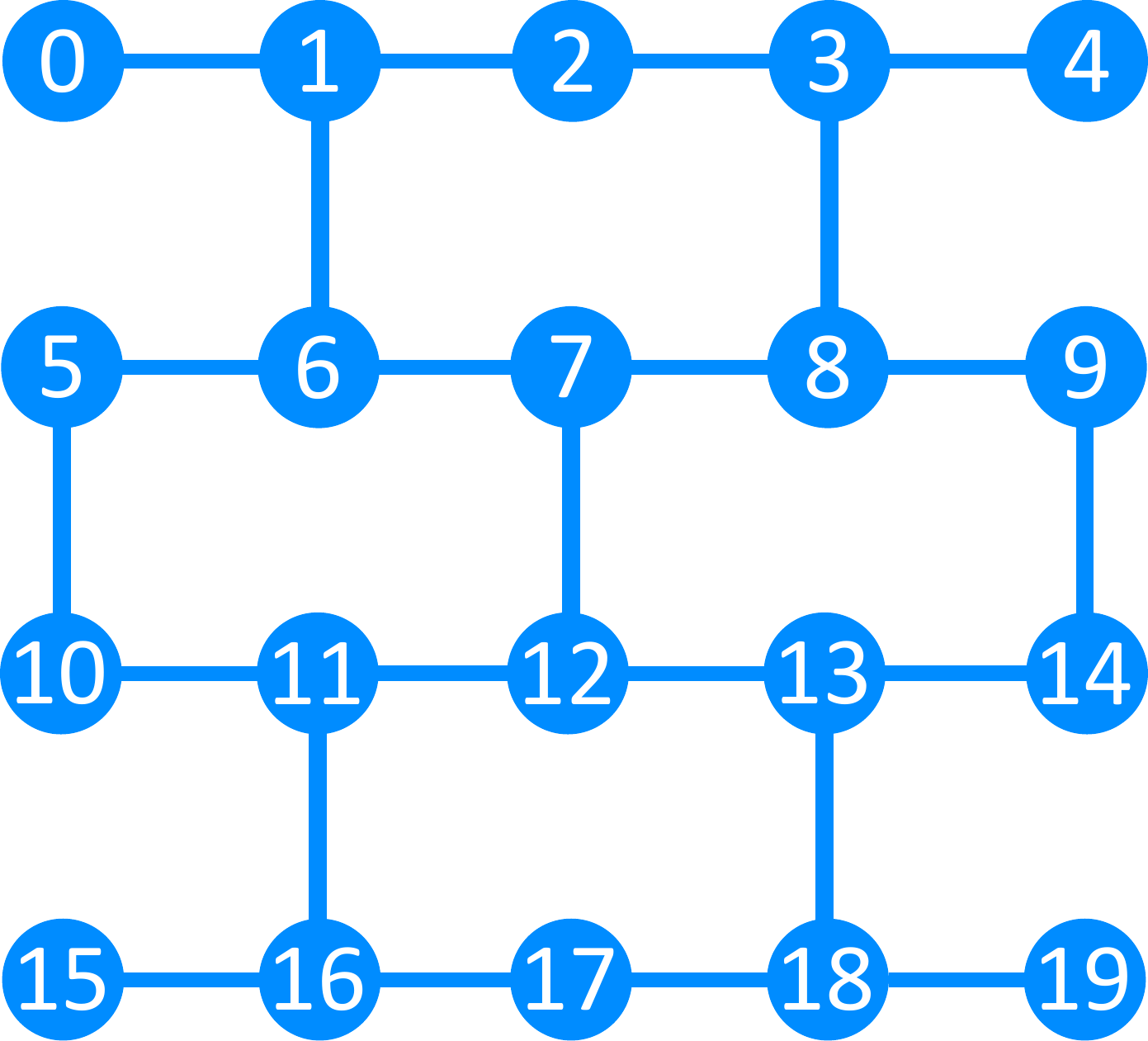}
		\subcaption{}
		\label{fig:coupling_maps_b}
		\end{minipage}
	\end{tabular}
\caption{\textbf{(a)} 20 qubits \textit{ibmq\_tokyo} and \textbf{(b)} \textit{ibmq\_almaden}~\cite{new_backends}.}
\label{fig:coupling_maps}
\end{figure}

\subsection{CNOT cascades}
\label{sec:cnot_cascade}

We are interested in the so called \textit{CNOT cascade}, shown in \REF{Fig.}{fig:cnot_cascade_decomposition_a}. This pattern plays a prominent role in several quantum algorithms such as the one used to produce GHZ states~\cite{GHZ1989,Deffner2017} as shown in \REF{Fig.}{fig:ghz}.

The coupling maps in \REF{Fig.}{fig:coupling_maps} prevent from placing all CNOT gates like in the ideal GHZ circuit, i.e., making a \textit{CNOT cascade} where $n - 1$ qubits control the \textit{n}th qubit. It is indeed possible to turn the ideal GHZ circuit into an equivalent one characterized by a unique sequence of CNOT gates. It would only be a slight change to the technique discussed in a previous work~\cite{Ferrari2018}.

However, this technique works only if the aim is to produce a GHZ state starting from a $|0\rangle^{\otimes n}$. This paper wants to focus on a more general case, where a CNOT cascade could appear at any point in the circuit and no assumption can be done on the state of the system.

\begin{figure}
\centering
	\begin{tabular}{ cc }
        \begin{minipage}{2.6cm}
        \centering
        \includegraphics[width=2.6cm]{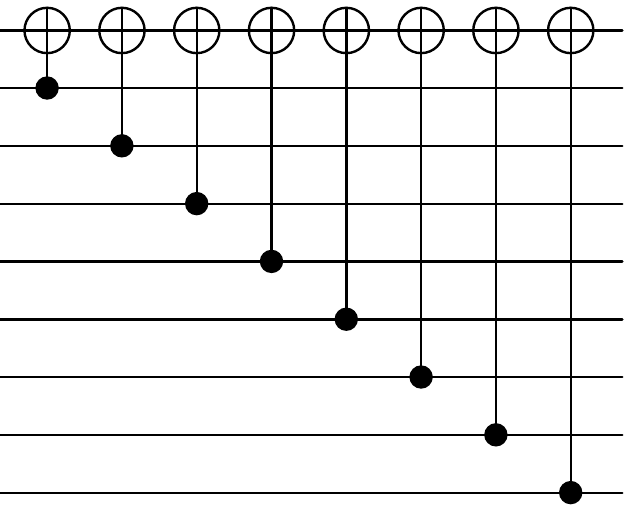}
        \subcaption{}
        \label{fig:cnot_cascade_decomposition_a}
        \end{minipage} & 
        \begin{minipage}{7cm}
        \centering
		\includegraphics[width=7cm]{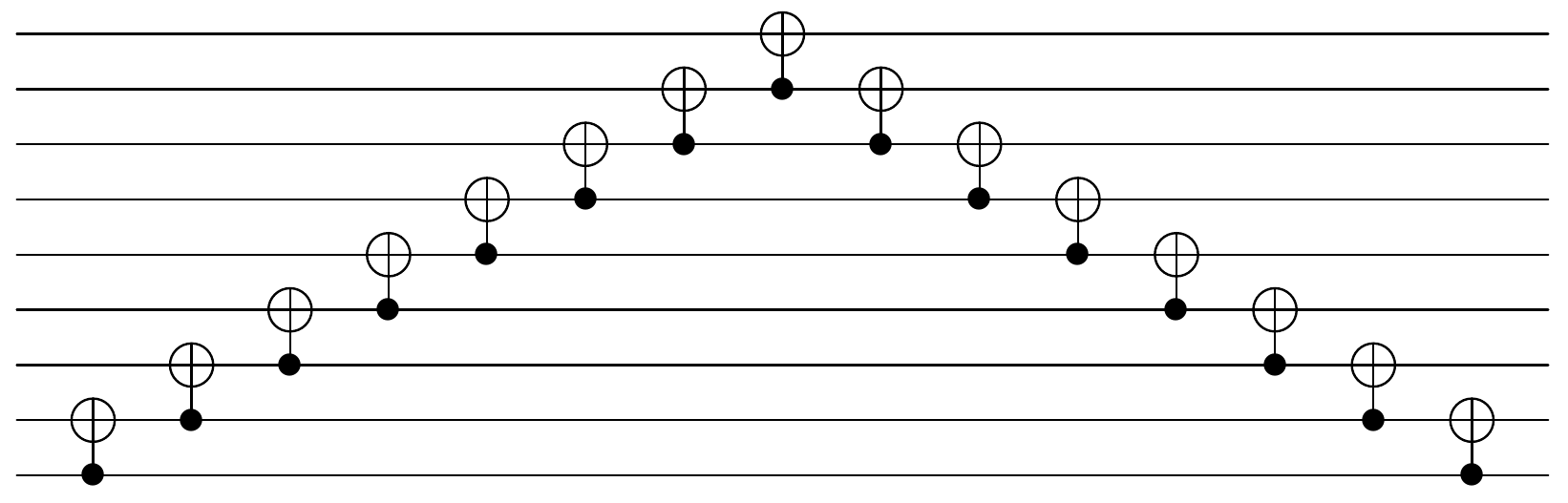}
		\subcaption{}
		\label{fig:cnot_cascade_decomposition_b}
		\end{minipage}\\
	\end{tabular}
\caption{\textbf{(a)} CNOT cascade. \textbf{(b)} Decomposition of a CNOT cascade}
\label{fig:cnot_cascade_decomposition}
\end{figure}

\begin{figure}[ht!]
\centering
	\begin{tabular}{ cc }
        \begin{minipage}{4.2cm}
        \centering
        \includegraphics[width=4.2cm]{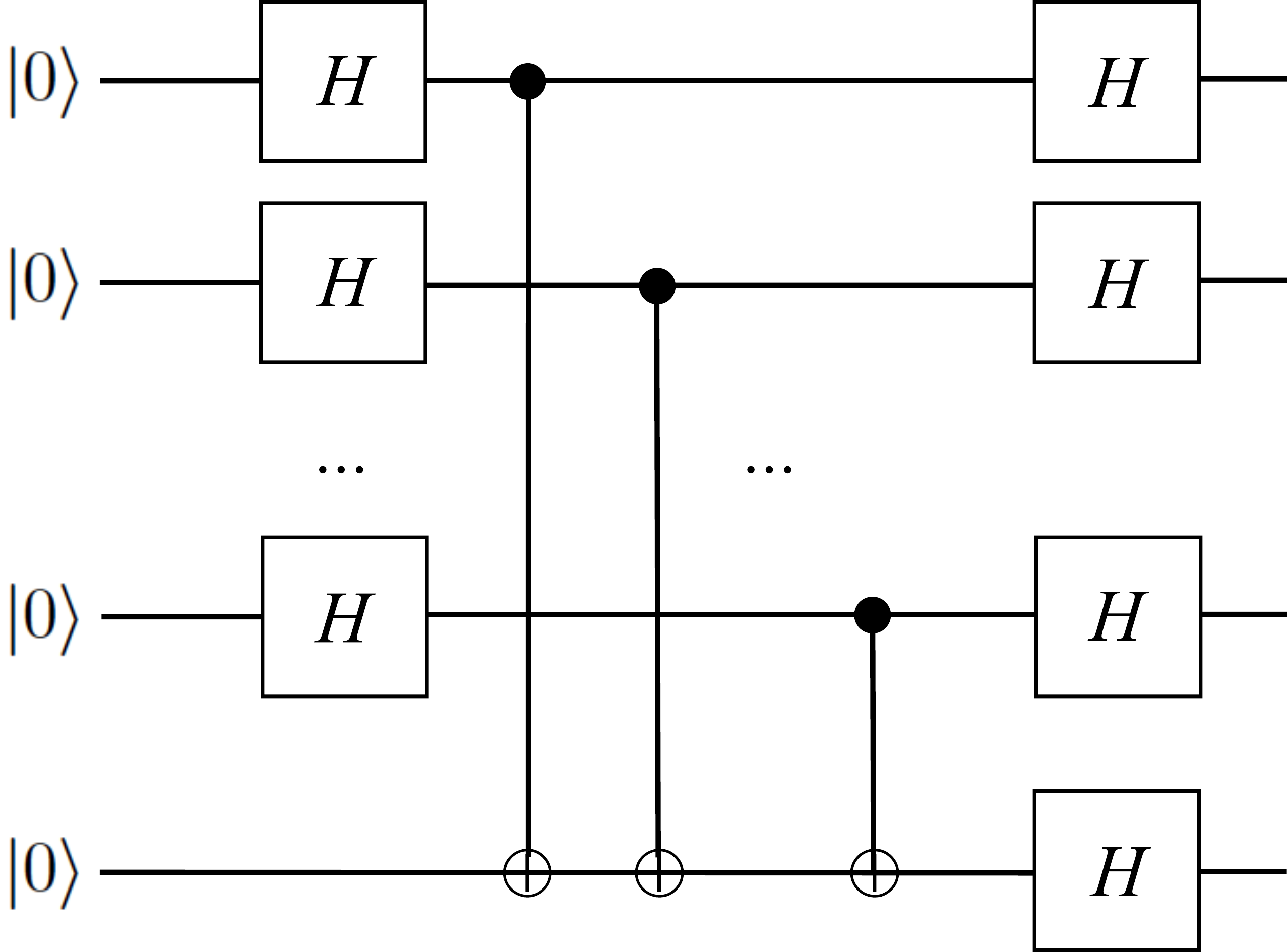}
        \end{minipage}
        & \boldmath$\frac{\KET{0}^{\otimes n}+\KET{1}^{\otimes n}}{\sqrt{2}}$\\
	\end{tabular}
\caption{GHZ circuit.}
\label{fig:ghz}
\end{figure}

A possible solution is to exploit the nearest neighbor decomposition for a uniformly controlled gate studied by Tucci~\cite{Tucci2004}. The only requirement for Tucci's decomposition is that qubits are to be arranged in a linear chain.

To realize the decomposition in \REF{Fig.}{fig:cnot_cascade_decomposition_b}, \REF{Algorithm}{alg:patterns} analyzes the circuit layer by layer and, when a CNOT gate is encountered, \REF{Algorithm}{alg:check_pattern1} checks if that CNOT is the first of a CNOT cascade. Each encountered CNOT cascade is replaced by its nearest-neighbor decomposition and, at the end, \REF{Algorithm}{alg:patterns} returns a new transformed circuit. Since \REF{Algorithm}{alg:patterns} just loops over all gates in a circuit, its complexity is $O(g)$ with $g$ being equal to the number of gates of the circuit.

\begin{algorithm}
	\footnotesize
	\caption{\textsc{Patterns($circuit$)}\newline
	\footnotesize
	\textbf{Input}: a quantum circuit $circuit$\newline
	\textbf{Output}: a new transformed circuit
}
	\label{alg:patterns}
	\begin{algorithmic}[1]
		\STATE $new\_circuit \gets \emptyset$
		\STATE to\_skip $\gets \emptyset$
		\STATE new\_layers $\gets [\emptyset \textbf{\space for \space}  0 \leq i < |circuit\_layers|]$
		
		\FOR{$i=0$ \TO $|circuit\_layers|$}
		    \IF{$i \neq 0$}
		        \FORALL{$g \in $ new\_layers[$i-1$]}
		            \STATE put \textit{g} into \textit{new\_circuit}
		        \ENDFOR
		    \ENDIF
		    
		    \FORALL{$gate \in circuit\_layers[i]$}
			\IF{$gate \not \in$ to\_skip}
				\IF{\textit{gate} is CNOT}
				\STATE transformed $\gets$ \textsc{CheckCascade($circuit\_layers$, $i$, new\_layers)}
				\IF{trasnformed $\neq \emptyset$}
					\STATE put transformed into to\_skip
					\STATE \textbf{continue}
				\ELSE
				    \STATE put $gate$ into to\_skip
					\STATE put \textit{gate} into \textit{new\_circuit}
				\ENDIF
			\ELSE
				\STATE put $gate$ into to\_skip
				\STATE put \textit{gate} into \textit{new\_circuit}
			\ENDIF
			\ENDIF
		\ENDFOR
		\ENDFOR
		\RETURN \textit{new\_circuit}
	\end{algorithmic}
\end{algorithm}

\REF{Algorithm}{alg:check_pattern1} analyzes the circuit starting from a CNOT gate; here \textit{before} and \textit{after} are sets of gates that can be applied before and after the decomposition. If a CNOT cascade is \textit{found}, the decomposition is applied between the before and after gates sets, otherwise an empty set is returned. In \REF{Algorithm}{alg:check_pattern1}, $gate_t$ is the target of $gate$ and if $gate$ is a CNOT, $gate_c$ is the control qubit of $gate$. As it is expected that cascades are no longer than the number of qubits in the circuit, the algorithm stops when, after $MAX=2m$ layers have been checked, no pattern is found. The time complexity of \REF{Algorithm}{alg:check_pattern1} is $O(m)$. Usually the number of gates $g$ is greater than $m$, thus, the complexity of \REF{Algorithm}{alg:patterns} is $O(g)$.

\begin{algorithm}
	\footnotesize
	\caption{\textsc{CheckCascade($layers$, $i$, $new\_layers$)}\newline
	\scriptsize
	\textbf{Input}: the list of layers in the circuit $layers$; the layer from where to start $i$; the list of $new\_layers$ to be added\newline
	\textbf{Output}: the list of gates to skip, $\emptyset$ if no cascade was found	
}
	\label{alg:check_pattern1}
	\begin{multicols}{2}
	\begin{algorithmic}[1]
		\STATE before $\gets \emptyset$, after $\gets \emptyset$, skip $\gets \emptyset$, off\_limits $\gets \emptyset$, control $\gets cnot_c$, target $\gets cnot_t$, used $\gets \emptyset$, found $\gets$ \textsf{false}, ctrls $\gets \emptyset$, put target into used, $i \gets 0$, $gate \gets circuit[0]$, $c \gets 0$, last $\gets i$
		\WHILE{$i < MAX$}
		\FORALL{$gate \in layers[i + c]$}
		    \IF{$gate \in$ skip \AND $gate_t =$ target}
		        \STATE $c \gets MAX$
		        \STATE \textbf{break}
		    \ELSE
		        \IF{$gate$ is a CNOT}
		            \IF{$gate_c =$ target}
		                \STATE $count \gets MAX$
		                \STATE \textbf{break}
		            \ELSIF{$gate_c \in$ off\_limits \OR $gate_t \in$ off\_limits}
		                \STATE put $gate_c$ into off\_limits
		                \STATE put $gate_c$ into used
		                \STATE put $gate_t$ into off\_limits
		                \STATE put $gate_t$ into used
		                \STATE \textbf{continue}
		            \ENDIF
		            \IF{$gate_t =$ target \AND $gate_c \not \in$ ctrls \AND $gate_c \not \in$ used}
		                \STATE put $gate_c$ into ctrls
		                \STATE put $gate_c$ into used
		                \STATE put $gate$ into skip
		            \ELSIF{$gate_t \neq$ target \AND $gate_c \neq$ target}
		                \IF{$gate_t \not \in$ used \AND $gate_c \not \in$ used \AND last $< c$}
		                    \STATE last $\gets i + c$
		                \ELSE
		                    \STATE put $gate_c$
		                    
		                    \hspace{\algorithmicindent} into off\_limits
		                    \STATE put $gate_c$
		                    
		                    \hspace{\algorithmicindent} into used
		                    \STATE put $gate_t$
		                    
		                    \hspace{\algorithmicindent} into off\_limits
		                    \STATE put $gate_t$
		                    
		                    \hspace{\algorithmicindent} into used
		                    \IF{last $> i + c - 1$}
		                    \STATE last $\gets i + c - 1$
		                    \ENDIF
		                \ENDIF
		            \ELSE
		                \STATE $count \gets MAX$
		                \STATE \textbf{break}
		            \ENDIF
		        \ELSE
		            \IF{$gate_t \in$ off\_limits}
		                \STATE \textbf{continue}
		            \ELSIF{$gate_t =$ target}
		                \STATE put $gate$ into after
		                \STATE put $gate$ into skip
		                \STATE $c \gets MAX$
		                \STATE \textbf{break}
		            \ELSIF{$gate_t \not \in$ used}
		                \STATE put $gate$ into before
		            \ELSE
		                \STATE put $gate$ into after
		            \ENDIF
		            \STATE put $gate$ into skip
		        \ENDIF
		    \ENDIF
		\ENDFOR
		\STATE $c \gets c + 1$
		\ENDWHILE
		\IF{$|$controls$|>1$}
			\FORALL{g $\in$ before}
			    \STATE put g into $new\_layers[$last$]$
			\ENDFOR
			\FORALL{$x \in$ reversed(ctrls) $\cup$ target}
				\STATE put $cnot_{x,x.next}$
				
				\hspace{\algorithmicindent} into $new\_layers[$last$]$
			\ENDFOR
			\FORALL{$y \in$ ctrls}
				\STATE put $cnot_{y,y.next}$
				
				\hspace{\algorithmicindent} into $new\_layers[$last$]$
			\ENDFOR
			\FORALL{g $\in$ after}
			    \STATE put g into $new\_layers[$last$]$
			\ENDFOR
			\RETURN skip
		\ENDIF
		\RETURN $\emptyset$
	\end{algorithmic}
	\end{multicols}
\end{algorithm}

We recall that the aim is to compile circuits characterized by repeated patterns. Let us look at the circuit in \REF{Fig.}{fig:cnot_cascade_sequences_a}, where multiple CNOT cascades are repeated one after the other, acting on different target qubits.
\REF{Algorithm}{alg:patterns} outputs the circuit in \REF{Fig.}{fig:cnot_cascade_sequences_b}, which despite being correct, has an increased depth.

\begin{figure}[h!]
\centering
	\begin{tabular}{ cc }
        \begin{minipage}{5cm}
        \centering
        \includegraphics[width=5cm]{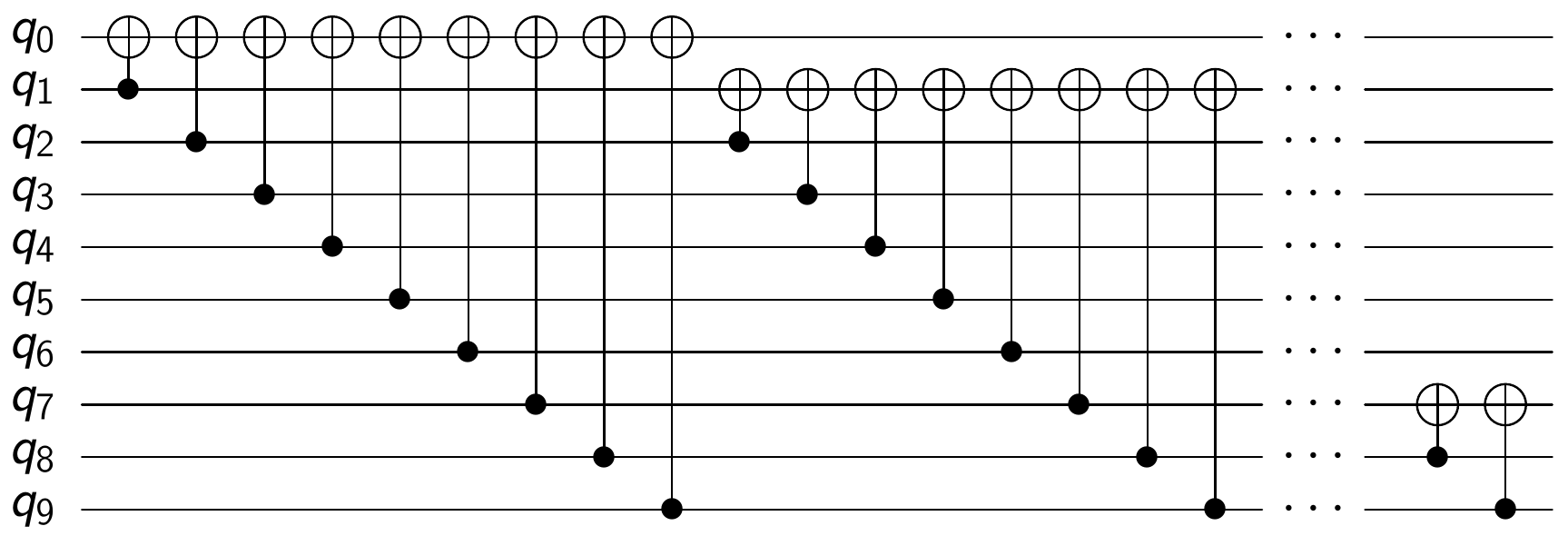}
        \subcaption{}
        \label{fig:cnot_cascade_sequences_a}
        \end{minipage} &\\
        \begin{minipage}{8.5cm}
        \centering
		\includegraphics[width=8.5cm]{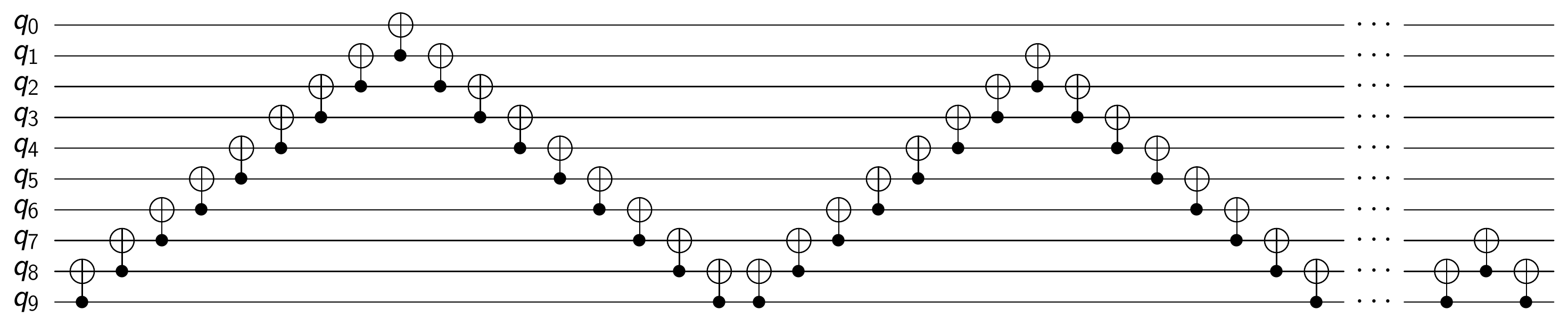}
		\subcaption{}
		\label{fig:cnot_cascade_sequences_b}
		\end{minipage} &
		\begin{minipage}{2.4cm}
		\centering
		\includegraphics[width=2.4cm]{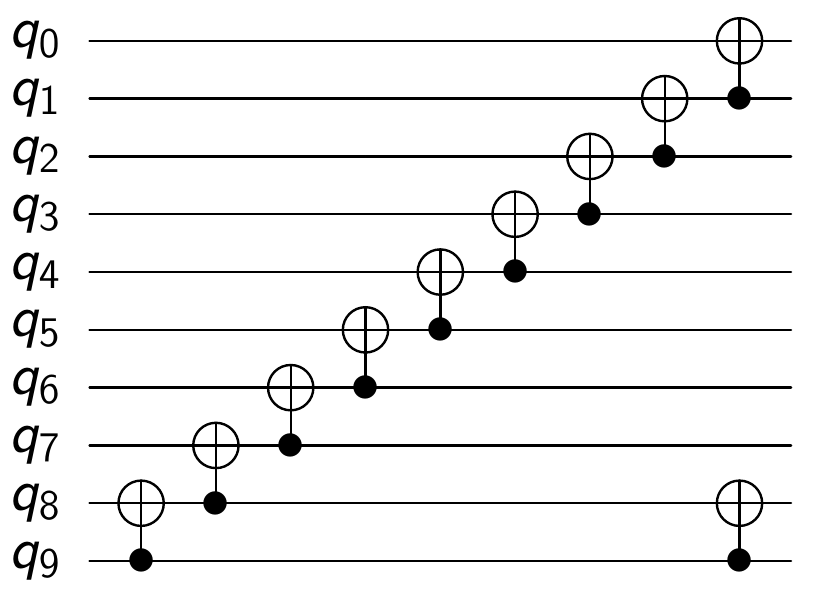}
		\subcaption{}
		\label{fig:cnot_cascade_sequences_c}
		\end{minipage}\\
	\end{tabular}
\caption{\textbf{(a)} Circuit with multiple CNOT cascades. \textbf{(b)} Circuit after nearest-neighbor decomposition. \textbf{(c)} Circuit after CNOT cancellation.}
\label{fig:cnot_cascade_sequences}
\end{figure}

Fortunately, if two consecutive CNOT gates act on the same control and target qubit, they elide each other, as a CNOT gate is the inverse gate of itself. \REF{Algorithm}{alg:cnot_cancellation} loops over the circuit to cancel double CNOT gates until no further cancellation can be done.

Since each \textit{layer} in the circuit has at most $m$ gates, in the worst case \REF{Algorithm}{alg:cnot_cancellation} needs $m$ iterations to cancel every CNOT couple. Its time complexity is $O(lm^2)$, where $l$ is the number of layers and usually $m \ll l$.

\begin{algorithm}
	\footnotesize
	\caption{\textsc{CnotCancellation($circuit$)}\newline
	\footnotesize
	\textbf{Input}: a quantum circuit $circuit$\newline
	\textbf{Output}: a new circuit without double CNOTs
}
	\label{alg:cnot_cancellation}
	\begin{algorithmic}[1]
		\STATE changed $\gets$ \textsf{true}
		\WHILE{changed is \textsf{true}}
			\STATE changed $\gets$ \textsf{false}
			\FORALL{$layer \in circuit$}
				\FORALL{$gate \in layer$}
					\IF{$gate$ is cnot}
						\IF{$cnot_{gate_c,gate_t} \in layer.next$}
							\STATE remove gate from \textit{layer}
							\STATE remove $cnot_{gate_c,gate_t}$ from \textit{layer.next}
							\STATE changed $\gets$ \textsf{true}
						\ENDIF
					\ENDIF
				\ENDFOR
			\ENDFOR
		\ENDWHILE
	\end{algorithmic}
\end{algorithm}

The result of such an optimization is shown in \REF{Fig.}{fig:cnot_cascade_sequences_c}. This new optimized circuit has a much more acceptable depth, even lower than the original circuit shown in \REF{Fig.}{fig:cnot_cascade_sequences_a}.

\subsection{Inverted CNOT cascades}
\label{sec:inverted_cascade}

\begin{figure}
	\centering
	\includegraphics[width=5.5cm]{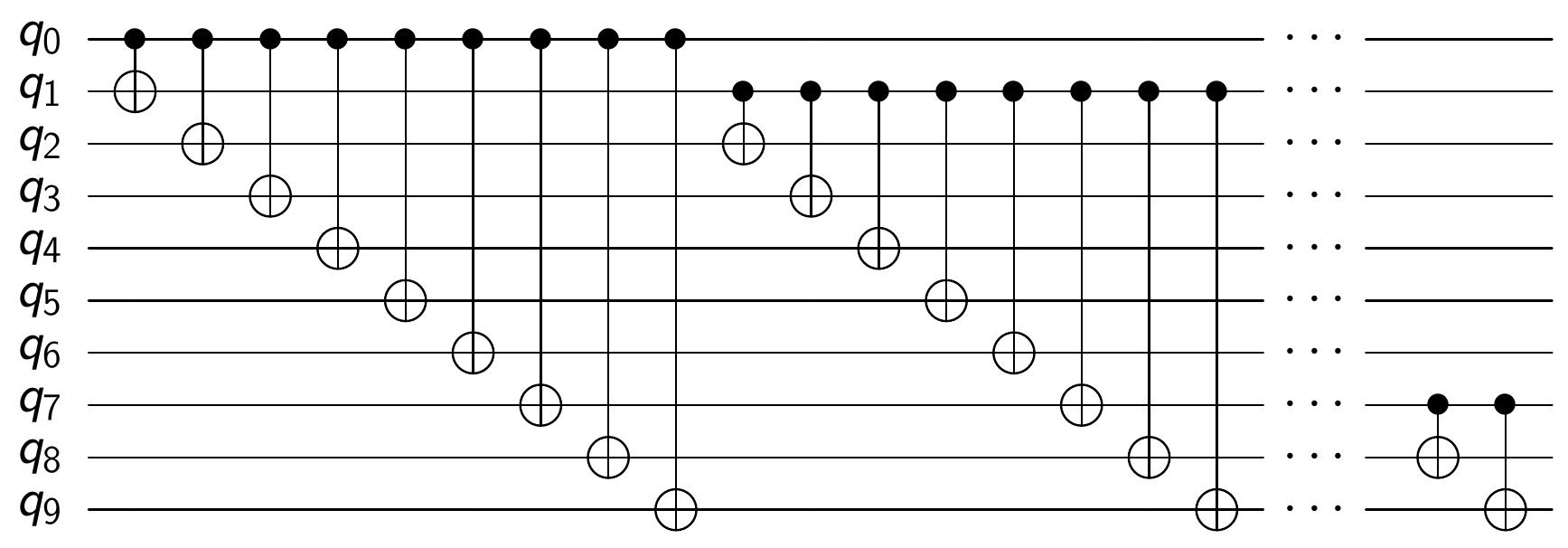}
	\caption{Circuit with multiple inverse CNOT cascades.}
	\label{fig:cnot_cascade_sequence_inverse}
\end{figure}

The pattern characterizing RyRz circuits, shown in \REF{Fig.}{fig:cnot_cascade_sequence_inverse}, is very similar to the one in \REF{Fig.}{fig:cnot_cascade_decomposition_a}. Indeed this pattern can be turned into the other one by inverting all of its CNOT gates by means of \textit{H} gates on both control and target qubits before and after the CNOT.
Of course adding \textit{H} gates to invert a CNOT would alter the circuit identity, therefore, instead of adding an \textit{H} gate, two \textit{H} gates are added so that they can negate each other's effects and leave the circuit identity untouched. The result of this operation is shown in \REF{Fig.}{fig:cnot_cascade_sequence_inversed_a}.

\begin{figure}
\centering
	\begin{tabular}{ cc }
        \multicolumn{2}{ c }{
        \begin{minipage}{11cm}
        \centering
        \includegraphics[width=11cm]{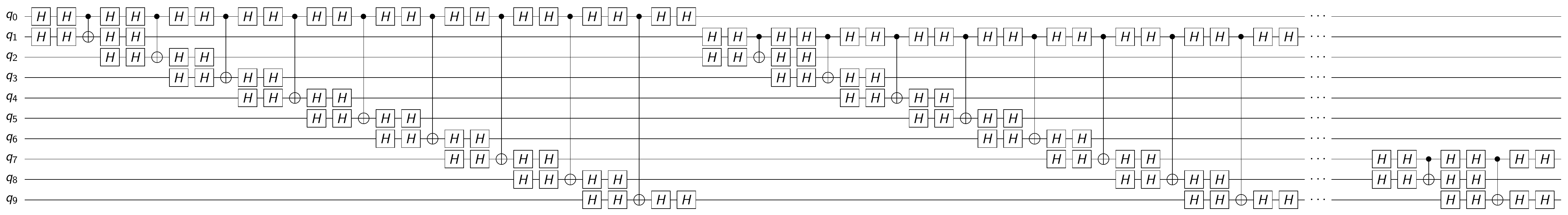}
        \subcaption{}
        \label{fig:cnot_cascade_sequence_inversed_a}
        \end{minipage}}\\
        &\\
        \begin{minipage}{8.5cm}
        \centering
		\includegraphics[width=8.5cm]{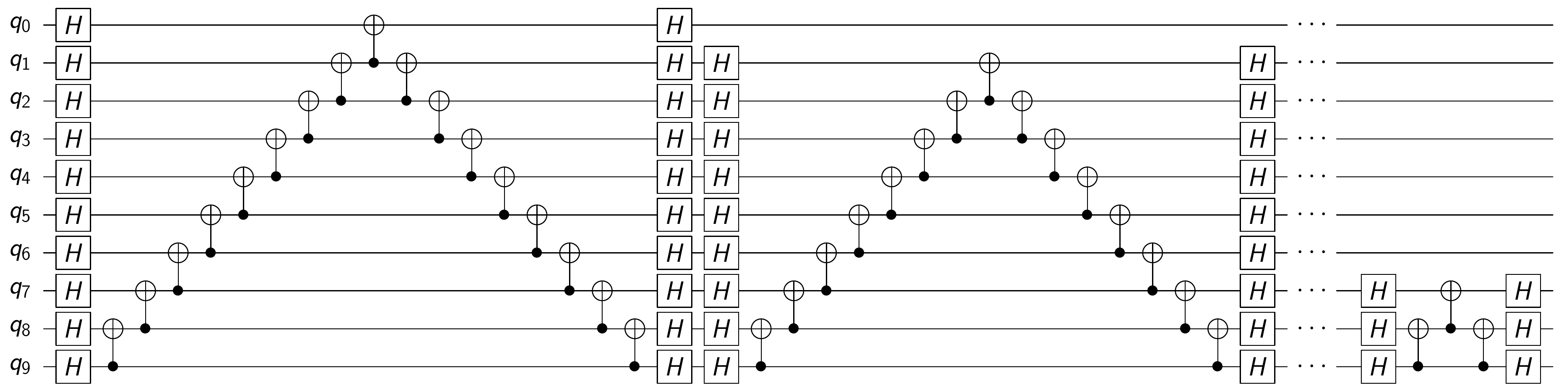}
		\subcaption{}
		\label{fig:cnot_cascade_sequence_inversed_b}
		\end{minipage} &
		\begin{minipage}{2cm}
		\centering
		\includegraphics[width=2cm]{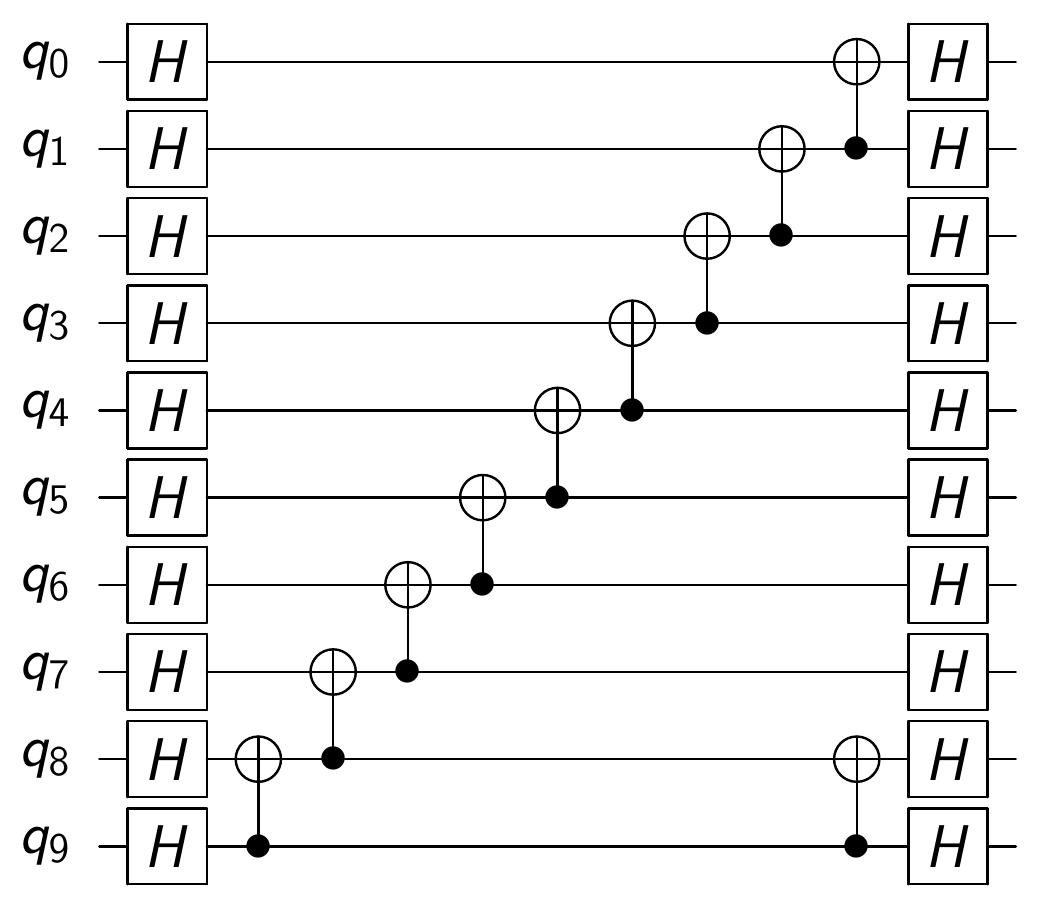}
		\subcaption{}
		\label{fig:cnot_cascade_sequence_inversed_c}
		\end{minipage}
	\end{tabular}
\caption{\textbf{(a)} Circuit with multiple inverse CNOT cascades after CNOT inversion. \textbf{(b)} Circuit with multiple inverse CNOT cascades after nearest-neighbor decomposition. \textbf{(c)} Circuit with multiple inverse CNOT cascades after gates cancellation.}
\label{fig:cnot_cascade_sequence_inversed}
\end{figure}

Using an algorithm similar to \REF{Algorithm}{alg:check_pattern1}, with the roles of control and target qubits exchanged, and a slight modification to \REF{Algorithm}{alg:patterns}, the circuit in \REF{Fig.}{fig:cnot_cascade_sequence_inversed_a} can be compiled obtaining the one in \REF{Fig.}{fig:cnot_cascade_sequence_inversed_b}. \REF{Algorithm}{alg:gate_cancellation} optimizes the circuit producing the one shown in \REF{Fig.}{fig:cnot_cascade_sequence_inversed_c}. Like in the previous case, the depth of the recompiled circuit is very close to the original.
The time complexity of \REF{Algorithm}{alg:gate_cancellation} is $O(lm^2)$.

\begin{algorithm}
	\footnotesize
	\caption{\textsc{GateCancellation($circuit$)}\newline
	\footnotesize
	\textbf{Input}: $circuit$ a quantum circuit\newline
	\textbf{Output}: a new circuit without double CNOTs and double \textit{H}
}
	\label{alg:gate_cancellation}
	\begin{algorithmic}[1]
		\STATE changed $\gets$ \TRUE
		\WHILE{changed is \textsf{true}}
			\STATE changed $\gets$ \textsf{false}
			\FORALL{$layer \in circuit$}
				\FORALL{$gate \in layer$}
					\IF{$gate$ is cnot}
						\IF{$cnot_{gate_c,gate_t} \in layer.next$}
							\STATE remove gate from \textit{layer}
							\STATE remove $cnot_{gate_c,gate_t}$ from \textit{layer.next}
							\STATE changed $\gets$ \textsf{true}
						\ENDIF
					\ELSIF{$gate$ is $H$}
						\IF{$gate \in layer.next$}
							\STATE remove $gate$ from \textit{layer}
							\STATE remove $gate$ from \textit{layer.next}
							\STATE changed $\gets$ \textsf{true}
						\ENDIF
					\ENDIF
				\ENDFOR
			\ENDFOR
		\ENDWHILE
	\end{algorithmic}
\end{algorithm}

\subsection{Nearest neighbor CNOT sequences}
\label{sec:nn_cnot}

In the previous section we showed how to transform inverted CNOT cascades into nearest-neighbor CNOT sequences. Clearly such a sequence of gates, where every qubit $q_i$ controls $q_{i+1}$, cannot be directly executed on the coupling map in \REF{Fig.}{fig:coupling_maps_b}, as $q_{i}$ is not always connected with $q_{i+1}$.

A possible solution is to find a path in the coupling map such that every qubit $q_i$, with $1<i<n-1$, with $n$ being the number of qubits in the device, has a connection with its nearest neighbors $q_{i-1}$ and $q_{i+1}$. This is related to the problem of finding an \textit{Hamiltonian path}, i.e., a path that visits each vertex of a graph exactly once. The Hamiltonian path problem is a special case of the \textit{Hamiltonian cycle} problem and is known to be NP-Complete, in fact it is an instance of the famous \textit{traveling salesman} problem \cite{Papadimitriou1998}.

Fortunately, one can take advantage of the features of the coupling map such as its regular structure and the fact that every qubit is identified by a number ranging from $0$ to $n$. As each undirected link can be seen as a couple of ingoing and outgoing links, the path obtained resembles a \textit{chain} and will be denoted as such, from now on.

\REF{Algorithm}{alg:chain} computes a chain in an undirected graph $\mathcal{G}$ starting from node $0$, where $\mathcal{C}$ is the chain initialized as empty, $\mathcal{S}$  is the set of nodes to be explored and $\mathcal{N}_x$ is the set of neighbors of node $x$. The algorithm loops over the nodes until the set of explored nodes $\mathcal{E}$ is equal to the set of nodes in $\mathcal{G}$. For every node added to $\mathcal{C}$, the \textsf{chain()} algorithm checks if the node's neighbors lead to a dead end, i.e., are isolated. If one neighbor is found to be isolated, it is added to the set of isolated nodes $\mathcal{I}$ and also to $\mathcal{E}$.

\begin{figure}
\centering
	\includegraphics[width=10cm]{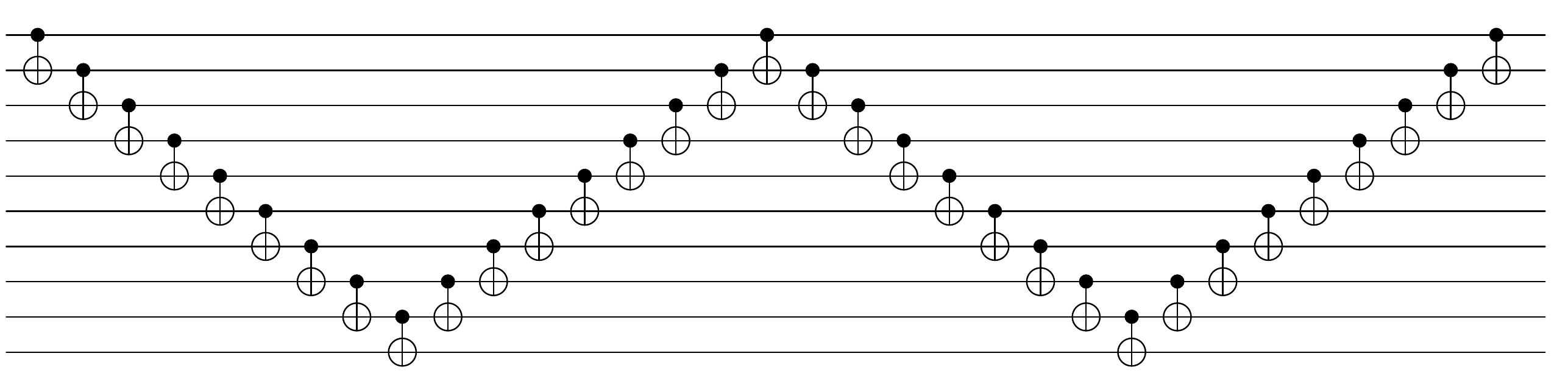}
	\caption{Circuit with nearest neighbor CNOT sequences.}
	\label{fig:nn_cnot_seq}
\end{figure}

\begin{algorithm}
	\footnotesize
	\caption{
    	\textsc{Chain($\mathcal{G}$, $n$)}\newline
    	\footnotesize
    	\textbf{Input}: undirected graph $\mathcal{G}$; number of qubits $n$ used by the circuit\newline
    	\textbf{Output}: a chain $C$ connecting at least $n$ nodes in $\mathcal{G}$
    }
	\label{alg:chain}
	\begin{algorithmic}[1]
	\STATE $\mathcal{C} \gets \emptyset$
	\STATE $\mathcal{S} \gets$ all nodes of $\mathcal{G}$
	\STATE put $0$ into $\mathcal{C}$
	\STATE $\mathcal{S} \gets \mathcal{S}/0$
	\STATE $\mathcal{E} \gets 0$
	\STATE $\mathcal{I} \gets \emptyset$
	\STATE $x = \mathcal{C}[|\mathcal{C}|-1]$
	\STATE $last\_back\_step \gets -1$
	\WHILE{$|\mathcal{E}|<|\mathcal{G}|$}
	    \STATE $\mathcal{N} \gets \mathcal{N}_x/E$
	    \IF{$|\mathcal{N}| \neq \emptyset$}
	        \IF{$x+1 \in \mathcal{N}_x$}
	            \STATE $x \gets x+1$
	        \ELSE
	            \STATE $x \gets min(\mathcal{N}_x)$
	        \ENDIF
	        \STATE put $x$ into $\mathcal{E}$
	        \STATE put $x$ into $\mathcal{C}$
	        \STATE $\mathcal{S} \gets \mathcal{S}/x$
	        \IF{$|\mathcal{E}|<|\mathcal{G}|-1$}
	            \STATE $\mathcal{N} \gets \emptyset$
	            \FORALL{$q \in \mathcal{N}_x$}
	            \IF{$q \not\in \mathcal{E}$}
	                \STATE remove=\textbf{true}
	                    \IF{$|N_q|=1$ \AND $|\mathcal{E}<|G|-1|$}
	                        \STATE put $q$ into $\mathcal{E}$
	                        \STATE $\mathcal{S} \gets \mathcal{S}/q$
	                        \STATE put $q$ into $\mathcal{I}$
	                        \STATE \textbf{continue}
	                    \ENDIF
	                    \FORALL{$r \in \mathcal{N}_q$}
	                        \IF{$r \not\in \mathcal{E}$ \AND $r = x$}
	                            \STATE remove=\textbf{false}
	                        \ENDIF
	                        \IF{remove=\textbf{true}}
	                            \STATE put $q$ into $\mathcal{E}$
	                            \STATE $\mathcal{S} \gets \mathcal{S}/q$
	                            \STATE put $q$ into $\mathcal{I}$
	                        \ENDIF
	                    \ENDFOR
	                \ENDIF
	           \ENDFOR
	        \ENDIF
	    \ELSE
	        \IF{$last\_back\_step \neq \mathcal{C}[|\mathcal{C}|-2]$ \AND $|\mathcal{G}|-|\mathcal{E}| > |current - \mathcal{S}[0]|$}
	            \STATE \textbf{break}
	        \ENDIF
	        \STATE put $x$ into $\mathcal{I}$
	        \STATE $\mathcal{C} \gets \mathcal{C}/x$
	        \STATE $x \gets \mathcal{C}[|\mathcal{C}|-1]$
	        \STATE $last\_back\_step \gets x$
	    \ENDIF
	\ENDWHILE
	\IF{$|\mathcal{C}| \geq n$}
	    \RETURN $\mathcal{C}$
	\ENDIF
	\STATE \textsc{CheckForIsolated($\mathcal{G}$, $\mathcal{C}$, $\mathcal{E}$, $\mathcal{I}$)}
	\STATE \textsc{ExpandChain($\mathcal{G}$, $\mathcal{C}$, $\mathcal{I}$, $n$)}
	\RETURN $\mathcal{C}$
	\end{algorithmic}
\end{algorithm}

After executing \REF{Algorithm}{alg:chain} on the coupling map in \REF{Fig.}{fig:coupling_maps_b}, the path obtained can be used to formulate an initial layout for the circuit in \REF{Fig.}{fig:nn_cnot_seq}, such that logical qubit $q_i$ corresponds to chain element $\mathcal{C}[i]$. \REF{Fig.}{fig:almaden_path} shows the path obtained in the coupling map highlighted in red.
Such an initial layout eliminates the need to use SWAP gates and produces a circuit with ideally no increase in depth.

\begin{figure}
	\centering
	\includegraphics[width=6cm]{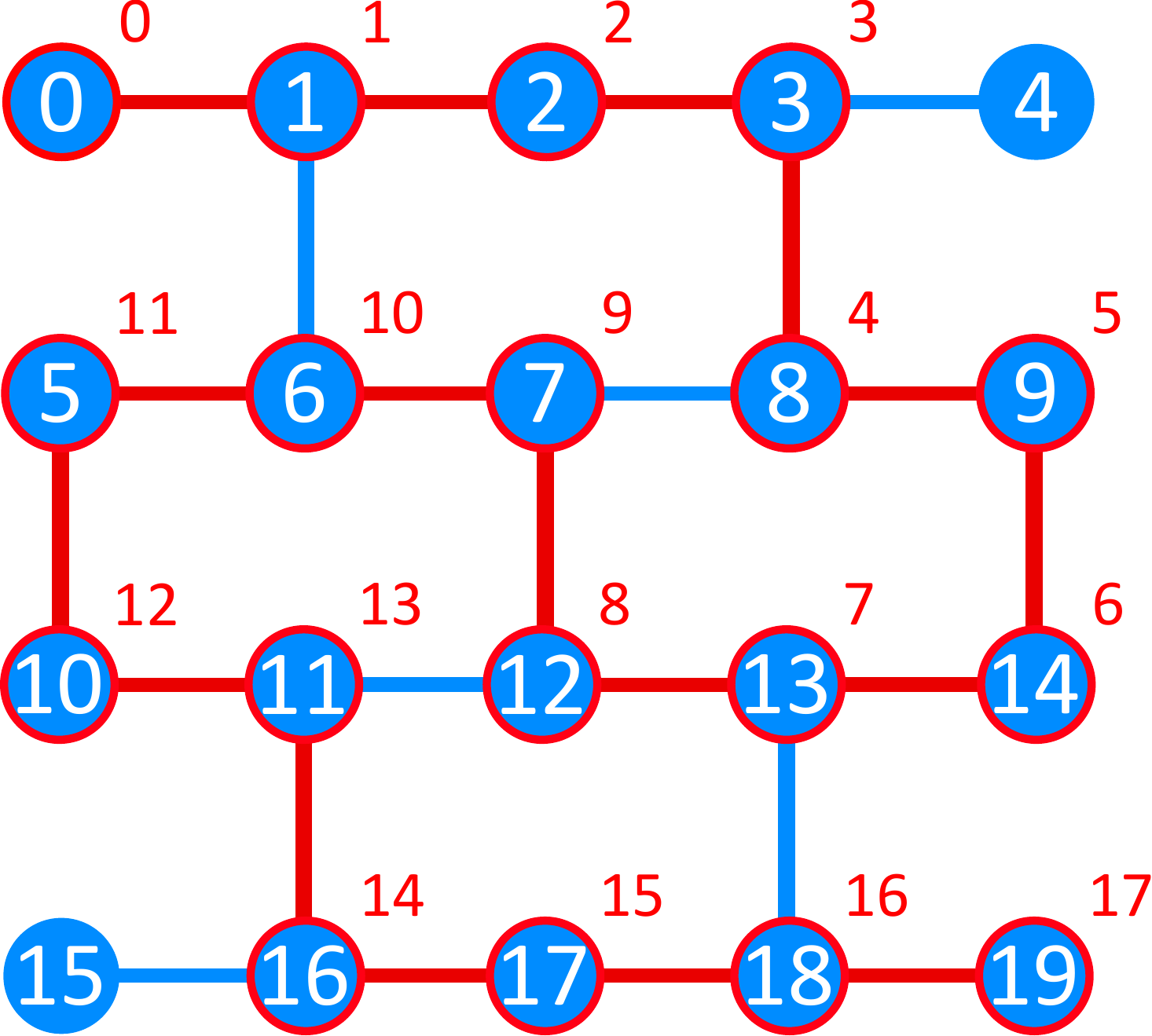}
	\caption{Qubit chain in \textit{ibmq\_almaden} highlighted in red.}
	\label{fig:almaden_path}
\end{figure}

\begin{algorithm}
	\footnotesize
	\caption{
    	\textsc{CheckForIsolated($\mathcal{G}$, $\mathcal{C}$, $\mathcal{E}$, $\mathcal{I}$)}\newline
    	\footnotesize
    	\textbf{Input}: an undirected graph $\mathcal{G}$; $\mathcal{C}$ a chain of nodes in $\mathcal{G}$; $\mathcal{E}$ nodes already explored; $\mathcal{I}$ nodes left outside $\mathcal{C}$ during exploration}
	\label{alg:isolated}
	\begin{algorithmic}[1]
	\FOR{$m=0$ \TO $|\mathcal{G}|-1$}
	    \IF{$m \not\in \mathcal{E}$ \AND $m \not\in \mathcal{I}$}
	        \FORALL{$i \in \mathcal{I}$}
	            \IF{$m \in \mathcal{N}_i$}
	                \STATE put $m$ into $\mathcal{I}$
	                \STATE put $m$ into $\mathcal{E}$
	                \STATE \textbf{break}
	            \ENDIF
	        \ENDFOR
	        \FORALL{$n \in \mathcal{N}_m$}
	            \IF{$n \in \mathcal{C}$}
	                \STATE put $m$ into $\mathcal{I}$
	                \STATE put $m$ into $\mathcal{E}$
	                \STATE \textbf{break}
	            \ENDIF
	        \ENDFOR
	    \ENDIF
	\ENDFOR
	\end{algorithmic}
\end{algorithm}

\begin{algorithm}
	\footnotesize
	\caption{
    	\textsc{ExpandChain($\mathcal{G}$, $\mathcal{C}$, $\mathcal{I}$, $n$)}\newline
    	\footnotesize
    	\textbf{Input}: an undirected graph $\mathcal{G}$; $\mathcal{C}$ a chain of nodes in $\mathcal{G}$; $\mathcal{I}$ nodes left outside $\mathcal{C}$ during exploration; $n$ the number of qubits used by the circuit}
	\label{alg:expand}
	\begin{algorithmic}[1]
	\STATE $r \gets (n-|\mathcal{C}|)$
	\WHILE{$r>0$}
	    \FORALL{$m \in \mathcal{I}$}
	        \STATE $x \gets min(\mathcal{N}_m \cap \mathcal{C})$
	        \IF{$x \neq \emptyset$}
	            \STATE put $m$ into $\mathcal{C}$ after $x$
	            \STATE $\mathcal{I} \gets \mathcal{I}/m$
	            \STATE $r \gets r-1$
	            \STATE \textbf{break}
	        \ENDIF
	    \ENDFOR
	\ENDWHILE
	\end{algorithmic}
\end{algorithm}

Nodes $4$ and $15$ are left outside the chain as it is already sufficiently long, to map the circuit in \REF{Fig.}{fig:nn_cnot_seq}, and adding these two extra qubits would require the use of SWAP gates during compilation. For a larger circuit, then those nodes will be inserted in the chain in a suitable position (nodes $4$ between $3$ and $8$, node $15$ between $16$ and $17$). Cycling through all $n$ nodes in the map, the time complexity of \REF{Algorithm}{alg:chain} is $O(n)$.

\section{Experimental results}
\label{sec:results}

We implemented the algorithms presented in Section \ref{sec:algorithms} in a quantum compiler written in Python language, denoted as PADQC\footnote{Source code: \url{https://github.com/qis-unipr/padqc}}.
In the PADQC framework, quantum circuits are represented by \textsf{QCircuit} objects, which are based on the well known formalism of Directed Acyclic Graphs (DAGs). In DAGs, nodes represent quantum gates and the edges connecting them correspond to qubits and bits. A \textsf{QCircuit} can then be easily converted into Open QASM~\cite{Cross2017open} and vice versa, allowing PADQC to interact with Qiskit.

\begin{table}[h!]
\Huge
\centering
{\resizebox{\textwidth}{!}{
    {\begin{tabular}{ | c | c | c | c | c | c | c | c | c | }
    \cline{3-9}
    \multicolumn{2}{ c }{}& \multicolumn{7}{ |c| }{CNOT Gate Count}\\
    \cline{3-9}
    \multicolumn{2}{c}{} & \multicolumn{1}{|c}{} & \multicolumn{3}{|c|}{\textit{ibmq\_tokyo}} & \multicolumn{3}{|c|}{\textit{ibmq\_almaden}}\\\hline
	Circuit Name & n & Initial & Qiskit(SABRE) & PADQC+Qiskit(SABRE) & t$|$ket$\rangle$ & Qiskit(SABRE) & PADQC+Qiskit(SABRE) & t$|$ket$\rangle$\\ \hline
	H2\_RYRZ & 4 & 30 & 30 & 21 & 30 & 60 & 21 & 70\\
	LiH\_RYRZ & 12 & 330 & 788 & 101 & 898 & 1256 & 101 & 1260\\
	H2O\_RYRZ & 14 & 455 & 1242 & 121 & 1301 & 1763 & 121 & 1738\\
	Random\_20q\_RYRZ & 20 & 950 & 2735 & 182 & 3053 & 3976 & 201 & 4146\\
\hline
\end{tabular}}}}
\caption{Circuit CNOT gate count of the compiled quantum chemistry circuits, considering the \textit{ibmq\_tokyo} and \textit{ibmq\_almaden} architectures.}
\label{tab:cnots}
\end{table}

\begin{table}[h!]
\Huge
\centering
{\resizebox{\textwidth}{!}{
    {\begin{tabular}{ | c | c | c | c | c | c | c | c | c | }
    \cline{3-9}
    \multicolumn{2}{ c }{}& \multicolumn{7}{ |c| }{CNOT Depth}\\
    \cline{3-9}
    \multicolumn{2}{c}{} & \multicolumn{1}{|c}{} & \multicolumn{3}{|c|}{\textit{ibmq\_tokyo}} & \multicolumn{3}{|c|}{\textit{ibmq\_almaden}}\\\hline
	Circuit Name & n & Initial & Qiskit(SABRE) & PADQC+Qiskit(SABRE) & t$|$ket$\rangle$ & Qiskit(SABRE) & PADQC+Qiskit(SABRE) & t$|$ket$\rangle$\\ \hline
	H2\_RYRZ & 4 & 21 & 21 & 21 & 21 & 60 & 21 & 70\\
	LiH\_RYRZ & 12 & 69 & 488 & 101 & 575 & 648 & 101 & 703\\
	H2O\_RYRZ & 14 & 81 & 675 & 121 & 813 & 846 & 121 & 875\\
	Random\_20q\_RYRZ & 20 & 117 & 1361 & 182 & 1648 & 1562 & 201 & 1629\\
\hline
\end{tabular}}}}
\caption{Circuit CNOT depth of the compiled quantum chemistry circuits, considering the \textit{ibmq\_tokyo} and \textit{ibmq\_almaden} architectures.}
\label{tab:cnots_depth}
\end{table}

Using an Intel Xeon E5-2683v4 2.1GHz with 50 GB of RAM, we benchmarked the quantum compiler with different quantum circuits. In particular, we considered a few quantum chemistry circuits for the RyRz heuristic~\cite{Kandala2017} wavefunction Ans\"atz  (\REF{Fig.}{fig:ryrz_circuits}), and an heterogeneous set of quantum circuits that has been used in most reference works~\cite{Li2019,Zulehner2019}. We assumed IBM Quantum hardware, namely \textit{ibmq\_tokyo} and \textit{ibmq\_almaden} architectures. They both have 20 qubits, but their coupling maps are quite different (Fig. \ref{fig:coupling_maps}).

We compared Qiskit(SABRE) with t$|$ket$\rangle$ and Qiskit(SABRE) preceded by PADQC transformations and mapping. Given that on NISQ devices multi-qubit operations tend to have error rates and execution times an order of magnitude worse than single-qubit ones~\cite{Arute2019}, the performance indicators used are CNOT gate gate count and CNOT depth of the circuit, i.e., the depth of the circuit taking into account only CNOT gates.
As shown in \REF{Table}{tab:cnots} and \REF{Table}{tab:cnots_depth}, the combination of PADQC and Qiskit(SABRE) outperforms both Qiskit(SABRE) alone and t$|$ket$\rangle$ on all RyRz circuits tested, regardless of the considered architecture.

We also tested PADQC on a larger set of circuits. We compiled a total of 190 circuits\footnote{Benchmark circuits QASM files: \url{https://github.com/qis-unipr/padqc/tree/master/benchmarks_qasm}}, taken from RevLib~\cite{RevLib}, Quipper~\cite{Quipper} and Scaffold~\cite{SacffCC}, plus the quantum chemistry ones. The results are summarized in \MULTIREF{Fig.}{fig:cnots}{fig:cnots_depth}, where the initial value of the considered metric is on the x-axis and the value for the compiled circuit on the y-axis.

\begin{figure*}
\centering
\begin{tabular}{ c }
    \begin{minipage}{14cm}
    	\centering
    	\includegraphics[width=14cm]{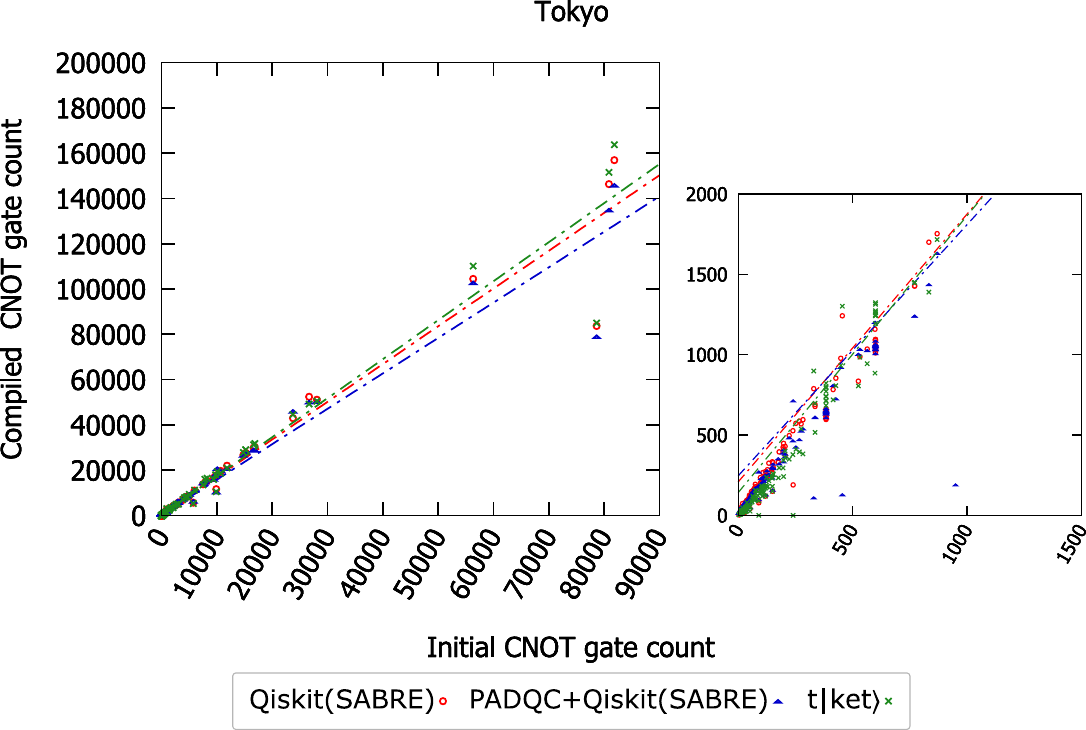}
        \label{fig:tokyo_cnots}
    \end{minipage}
    \\
    \begin{minipage}{14cm}
		\centering
		\includegraphics[width=14cm]{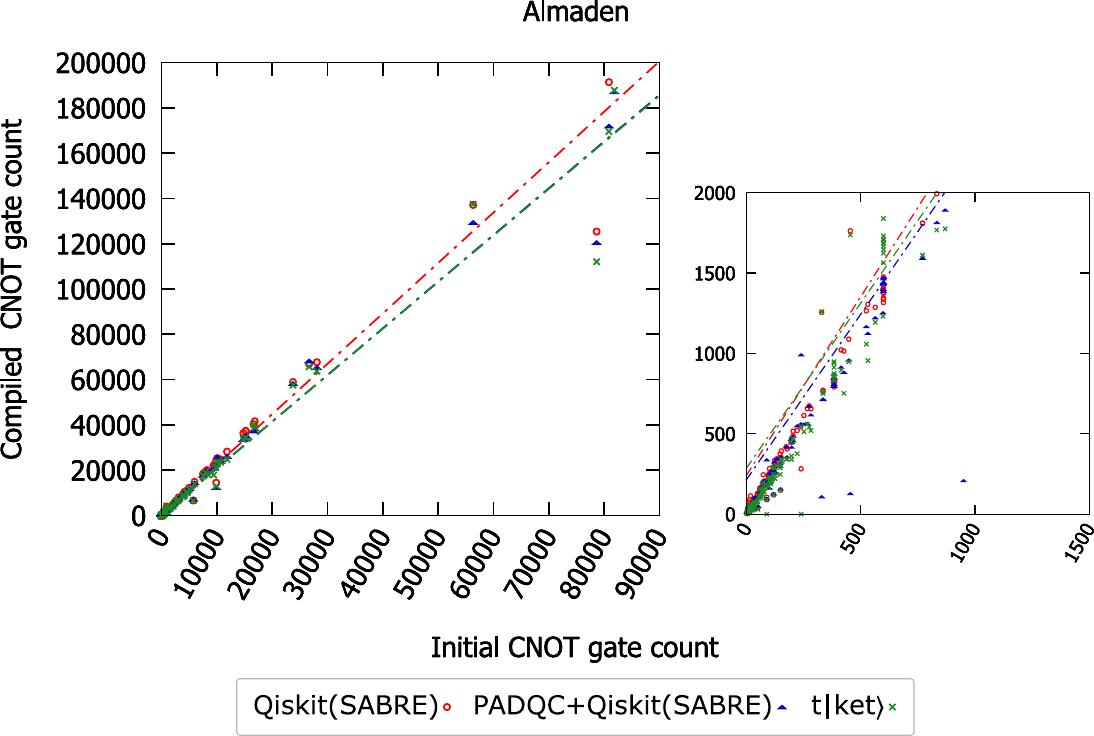}
		\label{fig:almaden_cnots}
	\end{minipage}\\
	\end{tabular}
    \caption{Comparing the CNOT gate count of the circuits compiled on \textit{ibmq\_tokyo} and \textit{ibmq\_almaden}.}
    \label{fig:cnots}
\end{figure*}

\begin{figure*}
\centering
\begin{tabular}{ c }
    \begin{minipage}{14cm}
    	\centering
    	\includegraphics[width=14cm]{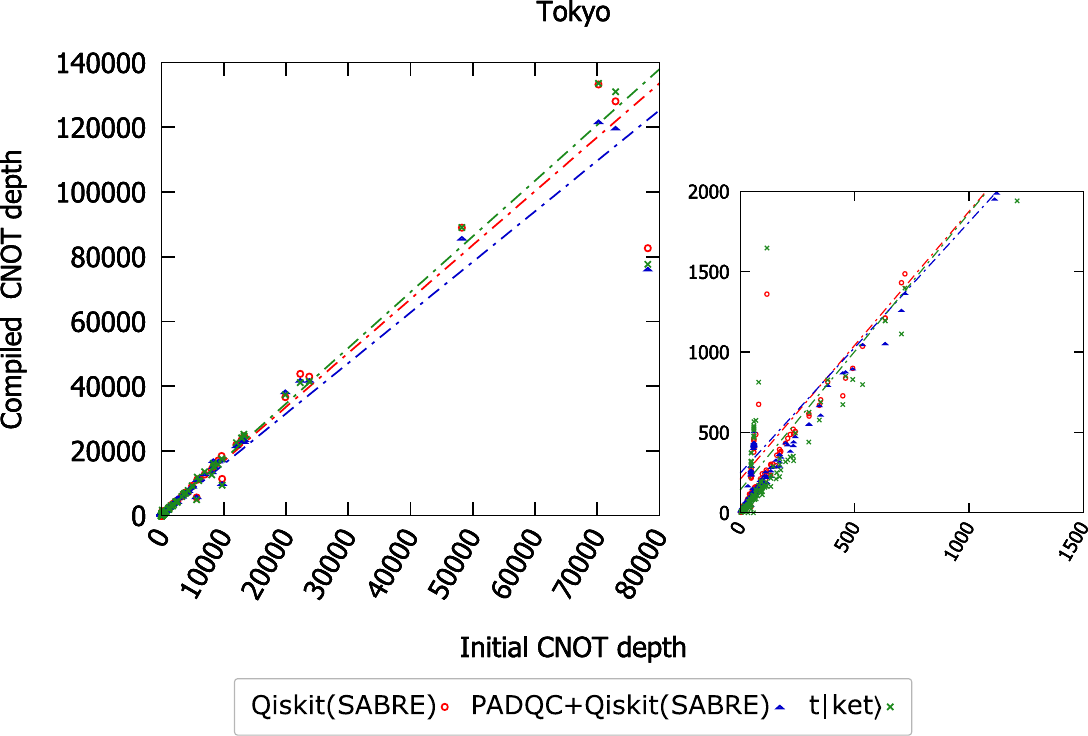}
        \label{fig:tokyo_cnots_depth}
    \end{minipage}
    \\
    \begin{minipage}{14cm}
		\centering
		\includegraphics[width=14cm]{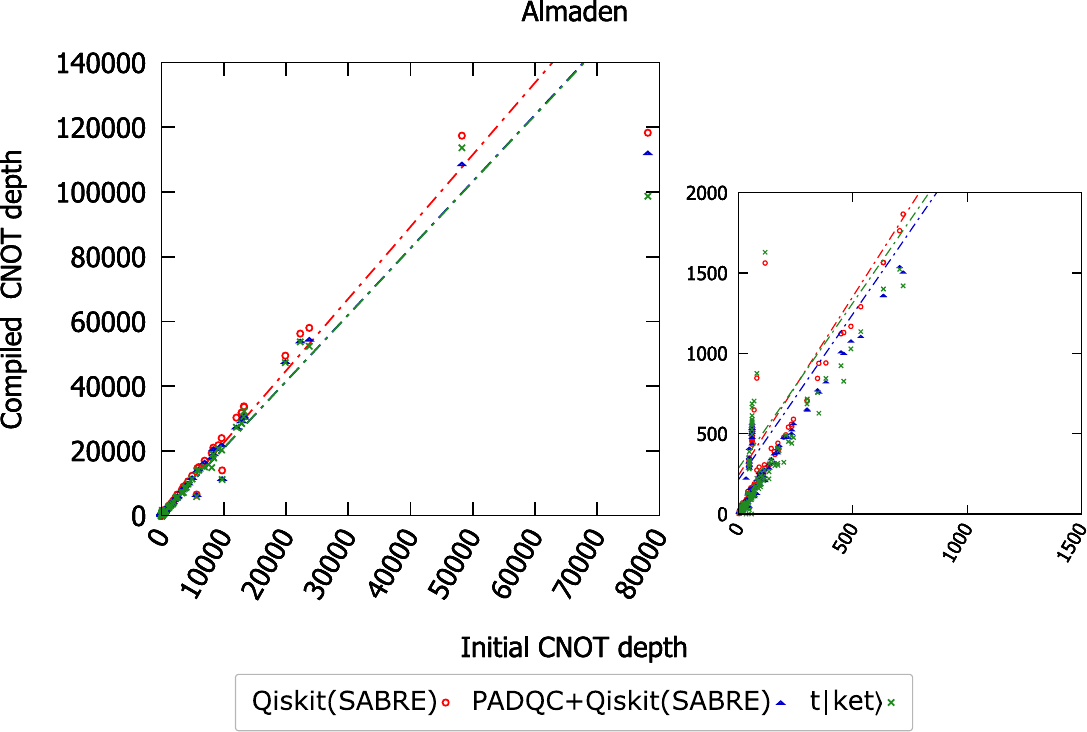}
		\label{fig:almaden_cnots_depth}
	\end{minipage}\\
	\end{tabular}
    \caption{Comparing the CNOT depth of the circuits compiled on \textit{ibmq\_tokyo} and \textit{ibmq\_almaden}.}
    \label{fig:cnots_depth}
\end{figure*}

When compiling on the \textit{ibmq\_tokyo} architecture, we can see, with the help of tendency lines calculated using the least squares method, that the combination of PADQC and Qiskit(SABRE) not only boosts Qiskit(SABRE) performance but can also compete with t$|$ket$\rangle$. As we switch to the less connected \textit{ibmq\_almaden} architecture, PADQC can still boost Qiskit(SABRE) performance while being up to par with t$|$ket$\rangle$.

\section{Conclusion}
\label{sec:conclusions}

In this paper, we proposed novel deterministic algorithms for compiling recurrent quantum circuit patterns in polynomial time. Starting from this set of algorithms, we have implemented PADQC, which has two main features. First, it identifies CNOT cascades and exploits useful circuit identities to transform them into CNOT nearest-neighbor sequences. Second, it finds an initial mapping that can comply with circuits characterized by recurrent CNOT nearest-neighbor sequences. Finally, we integrated PADQC with Qiskit’s SABRE swapping strategy and compilation routine.

We illustrated the results of the experimental evaluation of our integrated solution using different quantum programs and assuming IBM Quantum hardware. Among others, we compiled quantum circuits that are used to compute the ground state properties of molecular systems using the VQE method together with the RyRz heuristic wavefunction Ans\"atz.
PADQC+Qiskit(SABRE), in general, produces output programs that are on par with those produced by state-of-art tools, in terms of CNOT count and CNOT depth. In particular, our solution produces unmatched results on RyRz circuits.

In future work, we plan to expand PADQC to other patterns of interests and look for further circuit identities as the one found between nearest-neighbor CNOT sequences and CNOT cascade. These identities proved to be crucial in the quest for optimal quantum compilation. 
Moreover, we will investigate to possibility of integrating PADQC pattern transformation and mapping algorithms with alternative swapping strategies, which could produce an appreciable performance improvement.

While effective, CNOT count and CNOT depth are only pseudo-objectives. What really matters is the fidelity of a computation when run on actual quantum hardware.
We believe that the proposed compiler could be enhanced with noise related information (such as gate error rates and decoherence times) with the aim of finding a good trade-off between circuit depth and computation fidelity.

\section*{Acknowledgements}
This research benefited from the HPC (High Performance Computing) facility of the University of Parma, Italy.

\bibliographystyle{unsrt}
\bibliography{padqc.bib}

\begin{thebibliography}{10}

\bibitem{Feynman1982}
R.~P. {Feynman}.
\newblock Simulating physics with computers.
\newblock {\em International Journal of Theoretical Physics}, 21(6-7):467--488,
  1982.

\bibitem{Shor1994}
P.~W. {Shor}.
\newblock Algorithms for quantum computation: discrete logarithms and
  factoring.
\newblock In {\em Proceedings 35th Annual Symposium on Foundations of Computer
  Science}, pages 124--134, 1994.

\bibitem{Chiesa2018}
A.~{Chiesa}, F.~{Tacchino}, M.~{Grossi}, P.~{Santini}, I.~{Tavernelli},
  D.~{Gerace}, and S.~{Carretta}.
\newblock Quantum hardware simulating four-dimensional inelastic neutron
  scattering.
\newblock {\em Nature Physics}, 15:455--459, 2019.

\bibitem{Lloyd2014}
S.~{Lloyd}, M.~{Mohseni}, and P.~{Rebentrost}.
\newblock {Quantum principal component analysis}.
\newblock {\em Nature Physics}, 10(9):631--633, 2014.

\bibitem{Biamonte2017}
Jacob Biamonte, Peter Wittek, Nicola Pancotti, Patrick Rebentrost, Nathan
  Wiebe, and Seth Lloyd.
\newblock Quantum machine learning.
\newblock {\em Nature}, 549(7671), September 2017.

\bibitem{Havlicek2019}
Vojt\v{e}ch Havl\'i\v{c}ek, Antonio~D. C\'orcoles, Kristan Temme, Aram~W.
  Harrow, Abhinav Kandala, Jerry~M. Chow, and Jay~M. Gambetta.
\newblock Supervised learning with quantum-enhanced feature spaces.
\newblock {\em Nature}, 567(7747):209--212, March 2019.

\bibitem{Zoufal2019}
Christa Zoufal, Aurélien Lucchi, and Stefan Woerner.
\newblock Quantum {Generative} {Adversarial} {Networks} for learning and
  loading random distributions.
\newblock {\em npj Quantum Information}, 5(1):1--9, November 2019.

\bibitem{Cong2019}
Iris Cong, Soonwon Choi, and Mikhail~D. Lukin.
\newblock Quantum convolutional neural networks.
\newblock {\em Nature Physics}, 15(12), December 2019.

\bibitem{Tacchino2018}
F.~{Tacchino}, C.~{Macchiavello}, D.~{Gerace}, and D.~{Bajoni}.
\newblock An artificial neuron implemented on an actual quantum processor.
\newblock {\em NPJ Quantum Information}, 5:26:1--8, 2019.

\bibitem{Ekert1991}
A.~K. {Ekert}.
\newblock {Quantum cryptography based on Bell's theorem}.
\newblock {\em Phys. Rev. Lett.}, 67:661--663, 1991.

\bibitem{Portmann2014}
C.~{Portmann} and R.~{Renner}.
\newblock Cryptographic security of quantum key distribution.
\newblock arxiv:1409.3525, 2014.

\bibitem{Fitzsimons2017}
Joseph~F. Fitzsimons.
\newblock Private quantum computation: an introduction to blind quantum
  computing and related protocols.
\newblock {\em npj Quantum Information}, 3(1), June 2017.

\bibitem{Corcoles2020}
A.~D. {Córcoles}, A.~{Kandala}, A.~{Javadi-Abhari}, D.~T. {McClure}, A.~W.
  {Cross}, K.~{Temme}, P.~D. {Nation}, M.~{Steffen}, and J.~M. {Gambetta}.
\newblock Challenges and opportunities of near-term quantum computing systems.
\newblock {\em Proceedings of the IEEE}, 108(8):1338--1352, 2020.

\bibitem{Botea2018}
A.~Botea, A.~Kishimoto, and R.~Marinescu.
\newblock {On the Complexity of Quantum Circuit Compilation}.
\newblock In {\em The Eleventh International Symposium on Combinatorial Search
  (SOCS 2018)}, 2018.

\bibitem{Soeken2019}
M.~Soeken, G.~Meuli, B.~Schmitt, F.~Mozafari, H.~Riener, and G.~{De Micheli}.
\newblock Boolean satisfiability in quantum compilation.
\newblock {\em Phil. Trans. Royal Soc. A}, 378(2164):1--16, 2019.

\bibitem{Kliuchnikov2016}
V.~Kliuchnikov, D.~Maslov, and M.~Mosca.
\newblock {Practical Approximation of Single-Qubit Unitaries by Single-Qubit
  Quantum Clifford and T Circuits}.
\newblock {\em {IEEE Transactions on Computers}}, 65(1):161--172, 2016.

\bibitem{MunozCoreas2019}
E.~Munoz-Coreas and H.~Thapliyal.
\newblock {Quantum circuit design of a T-count optimized integer multiplier}.
\newblock {\em {IEEE Transactions on Computers}}, 68(5):729--739, 2019.

\bibitem{Kandala2017}
Abhinav Kandala, Antonio Mezzacapo, Kristan Temme, Maika Takita, Markus Brink,
  Jerry~M. Chow, and Jay~M. Gambetta.
\newblock Hardware-efficient variational quantum eigensolver for small
  molecules and quantum magnets.
\newblock {\em Nature}, 549(7671):242--246, September 2017.

\bibitem{Li2019}
Gushu Li, Yufei Ding, and Yuan Xie.
\newblock Tackling the qubit mapping problem for nisq-era quantum devices.
\newblock In {\em Proceedings of the Twenty-Fourth International Conference on
  Architectural Support for Programming Languages and Operating Systems},
  ASPLOS ’19, page 1001–1014, 2019.

\bibitem{QiskitSDK_short}
H{\'e}ctor Abraham et~al.
\newblock Qiskit: An open-source framework for quantum computing, 2019.

\bibitem{Sivarajah2020}
Seyon Sivarajah, Silas Dilkes, Alexander Cowtan, Will Simmons, Alec Edgington,
  and Ross Duncan.
\newblock t$|$ket$\rangle$: a retargetable compiler for {NISQ} devices.
\newblock {\em Quantum Science and Technology}, 6(1):014003, 2020.

\bibitem{Zulehner2019}
Alwin Zulehner, Alexandru Paler, and Robert Wille.
\newblock An efficient methodology for mapping quantum circuits to the {IBM}
  {QX} architectures.
\newblock {\em {IEEE} Trans. on {CAD} of Integrated Circuits and Systems},
  38(7):1226--1236, 2019.

\bibitem{RussellNorvig2020}
S.~{Russell} and P.~{Norvig}.
\newblock {\em Artificial Intelligence: A Modern Approach 4th Edition}.
\newblock Pearson, 2020.

\bibitem{Peruzzo2014}
A.~Peruzzo, J.~McClean, P.~Shadbolt, M.-H. Yung, X.-Q. Zhou, P.~J. Love,
  A.~{Aspuru-Guzik}, and J.~L. O'Brien.
\newblock A variational eigenvalue solver on a photonic quantum processor.
\newblock {\em {Nature Communications}}, 5(4213):1--7, 2014.

\bibitem{Barkoutsos2018}
Panagiotis~Kl. Barkoutsos, Jerome~F. Gonthier, Igor Sokolov, Nikolaj Moll, Gian
  Salis, Andreas Fuhrer, Marc Ganzhorn, Daniel~J. Egger, Matthias Troyer,
  Antonio Mezzacapo, Stefan Filipp, and Ivano Tavernelli.
\newblock {Quantum algorithms for electronic structure calculations:
  Particle-hole Hamiltonian and optimized wave-function expansions}.
\newblock {\em Phys. Rev. A}, 98:022322, Aug 2018.

\bibitem{new_backends}
IBM.
\newblock Quantum computation center opens.

\bibitem{GHZ1989}
D.~M. {Greenberger}, M.~A. {Horne}, and A.~{Zeilinger}.
\newblock Going beyond bell's theorem.
\newblock In M.~{Kafatos}, editor, {\em Bell's Theorem, Quantum Theory, and
  Conceptions of the Universe}, pages 69--72. Kluwer Academic Publishers, 1989.

\bibitem{Deffner2017}
S.~{Deffner}.
\newblock {Demonstration of entanglement assisted invariance on IBM's quantum
  experience}.
\newblock {\em Heliyon}, 3(11), 2017.

\bibitem{Ferrari2018}
D.~{Ferrari} and M.~{Amoretti}.
\newblock Efficient and effective quantum compiling for entanglement-based
  machine learning on ibm q devices.
\newblock {\em International Journal of Quantum Information}, 16(08), 2018.

\bibitem{Tucci2004}
Robert~R. Tucci.
\newblock {QC} {Paulinesia}.
\newblock arxiv:0407215, July 2004.

\bibitem{Papadimitriou1998}
C.~H. {Papadimitriou} and K.~{Steiglitz}.
\newblock {\em Combinatorial Optimization: Algorithms and Complexity}.
\newblock Dover Books on Computer Science. Dover Publications, 1998.

\bibitem{Cross2017open}
Andrew~W. Cross, Lev~S. Bishop, John~A. Smolin, and Jay~M. Gambetta.
\newblock {Open Quantum Assembly Language}.
\newblock arxiv:1707.03429, 2017.

\bibitem{Arute2019}
Frank Arute, Kunal Arya, Ryan Babbush, Dave Bacon, et~al.
\newblock Quantum supremacy using a programmable superconducting processor.
\newblock {\em Nature}, 574(7779):505--510, Oct 2019.

\bibitem{RevLib}
R.~{Wille}, D.~{Große}, L.~{Teuber}, G.~W. {Dueck}, and R.~{Drechsler}.
\newblock Revlib: An online resource for reversible functions and reversible
  circuits.
\newblock In {\em 38th International Symposium on Multiple Valued Logic (ismvl
  2008)}, pages 220--225, 2008.
\newblock {RevLib} is available at http://www.revlib.org.

\bibitem{Quipper}
Alexander~S. Green, Peter~FqiskiLeFanu Lumsdaine, Neil~J. Ross, Peter Selinger,
  and Beno\^{\i}t Valiron.
\newblock Quipper: A scalable quantum programming language.
\newblock {\em SIGPLAN Not.}, 48(6):333–342, June 2013.

\bibitem{SacffCC}
Ali JavadiAbhari, Shruti Patil, Daniel Kudrow, Jeff Heckey, Alexey Lvov,
  Frederic~T. Chong, and Margaret Martonosi.
\newblock Scaffcc: Scalable compilation and analysis of quantum programs.
\newblock {\em Parallel Computing}, 45:2 -- 17, 2015.
\newblock Computing Frontiers 2014: Best Papers.

\end{thebibliography}

\end{document}